\documentclass{aa}

\usepackage{graphicx}
\usepackage{color}
\usepackage{amsbsy}
\usepackage{mathrsfs}
\usepackage{mathpazo,bm}
\usepackage{amsmath}
\usepackage{multirow}
\usepackage{wasysym} 

\newlength{\hfwidth}
\newlength{\hfwidthsingle}
\newcommand{\pderiv}[2]{\frac{\partial #1}{\partial #2}}

\newcommand{\aderiv}[1]{\frac{{\mathcal D} #1}{{\mathcal D}t}}

\newcommand{\Rey}{\mathrm{Re}}

\newcommand{\Ro}{\mathrm{Ro}}
\newcommand{\Ri}{\mathrm{Ri}}
\newcommand{\ttimes}[1]{10^{#1}}
\newcommand{\xtimes}[2]{#1 \times 10^{#2}}

\newcommand{\vt}[1]{\mathbf{#1}}       
\renewcommand{\v}[1]{{\boldsymbol #1}} 

\newcommand{\cv}{c_{_V}}

\newcommand{\del}{\v{\nabla}}
\newcommand{\grad}{\del}
\newcommand{\Div}{\del\cdot}

\newcommand{\curl}{\del\times}
\newcommand{\Laplace}{\nabla^2}

\newcommand{\hatz}{\hat{\v{z}}}
\newcommand{\hatx}{\hat{\v{x}}}
\newcommand{\haty}{\hat{\v{y}}}
\newcommand{\va}{v_{_{\rm A}}}

\newcommand{\ksi}{\xi} 

\newcommand{\cte}{\small \rm {const}}

\newcommand{\Eq}[1]{Eq. (\ref{#1})}
\newcommand{\Eqs}[2]{Eqs. (\ref{#1}) and~(\ref{#2})}

\newcommand{\eq}[1]{\Eq{#1}}

\newcommand{\App}[1]{Appendix~\ref{#1}}
\newcommand{\app}[1]{\App{#1}}

\newcommand{\Fig}[1]{Fig.~\ref{#1}}
\newcommand{\fig}[1]{\Fig{#1}}
\newcommand{\Figs}[2]{Figs.~\ref{#1} and \ref{#2}}
\newcommand{\Figss}[3]{Figs.~\ref{#1},~\ref{#2}, and \ref{#3}}
\newcommand{\Figsss}[4]{Figs.~\ref{#1},~\ref{#2},~\ref{#3}, and \ref{#4}}
\newcommand{\Figssss}[5]{Figs.~\ref{#1},~\ref{#2},~\ref{#3},~\ref{#4}, and \ref{#5}}

\newcommand{\Tab}[1]{Table \ref{#1}}

\newcommand{\sect}[1]{Sect.~\ref{#1}}

\definecolor{brown}{rgb}{0.42,0.24,0.07}
\definecolor{darkgreen}{rgb}{0.0,0.6,0.00}
\definecolor{purple}{rgb}{0.7,0.0,0.7}
\definecolor{black}{rgb}{0.0,0.0,0.0}

\def\white#1{\textcolor{white}{#1}}
\def\black#1{\textcolor{black}{#1}}

\begin{document}

\title{The baroclinic instability in the context of layered accretion.} 
\subtitle{Self-sustained vortices and their magnetic stability in local compressible unstratified models of protoplanetary disks}

\author{W. Lyra\inst{1,2}, \& H. Klahr\inst{1}} 

\offprints{wlyra@amnh.org}

\institute{Max-Planck-Institut f\"ur Astronomie, K\"onigstuhl 17, 69117 Heidelberg, Germany
\and 
American Museum of Natural History, Department of Astrophysics, Central Park West at 79th Street, New York, NY, 10024-5192, USA}
\date{Received ; Accepted}

\authorrunning{Lyra \& Klahr}
\titlerunning{Baroclinic instability in magnetized protoplanetary disks}

\abstract
{Turbulence and angular momentum transport in accretion disks remains a topic 
of debate. With the realization that dead zones are robust features of 
protoplanetary disks, the search for hydrodynamical sources of turbulence 
continues. A possible source is the baroclinic instability (BI), which has been 
shown to exist in unmagnetized non-barotropic disks.}
{We aim to verify the existence of the baroclinic instability in 3D magnetized disks, 
as well as its interplay with other instabilities, namely the magneto-rotational 
instability (MRI) and the magneto-elliptical instability.}
{We performed local simulations of non-isothermal accretion 
disks with the {\sc Pencil Code}. The entropy gradient that generates the baroclinic 
instability is linearized and included in the momentum and energy equations in the 
shearing box approximation. The model is compressible, so excitation of spiral 
density waves is allowed and angular momentum transport can be measured.}
{We find that the vortices generated and sustained by the 
baroclinic instability in the purely hydrodynamical regime do not survive when 
magnetic fields are included. The MRI by far supersedes the BI in growth rate 
and strength at saturation.
The resulting turbulence is virtually identical to an MRI-only scenario.
We measured the intrinsic vorticity profile of the vortex, finding 
little radial variation in the vortex core. Nevertheless, the core is disrupted by 
an MHD instability, which we identify with the magneto-elliptic instability. This instability 
has nearly the same range of unstable wavelengths as the MRI, but has higher growth rates. 
In fact, we identify the MRI as a limiting case of the magneto-elliptic 
instability, \black{when the vortex aspect ratio tends to infinity (pure shear flow). 
We isolated its effect on the vortex, finding that a strong but 
unstable vertical magnetic field leads to channel flows inside the vortex, which 
stretch it apart. When the field is decreased or resistivity is used, we 
find that the vortex survives until the MRI develops in the box. The vortex is then 
destroyed by the strain of the surrounding turbulence. Constant azimuthal fields 
and zero net flux fields also lead to vortex destruction. Resistivity quenches both 
instabilities when the magnetic Reynolds number of the longest vertical wavelength of 
the box is near unity.}}
{We conclude that vortex excitation and self-sustenance by the baroclinic instability 
in protoplanetary disks is viable only in low ionization, i.e., the dead zone. Our results 
are thus in accordance with the layered accretion paradigm. A baroclinicly 
unstable dead zone should be characterized by the presence of large-scale vortices whose 
cores are elliptically unstable, yet sustained by the baroclinic feedback. 
Since magnetic fields destroy the vortices and the MRI outweighs the BI, the active 
layers are unmodified.}

\maketitle

\section{Introduction}
\label{sect:Introduction}

Turbulence is the preferred mechanism for enabling accretion in circumstellar 
disks, and the magneto-rotational instability (MRI, Balbus \& Hawley 1991) 
is the preferred route to turbulence. However, the MRI requires sufficient 
ionization since the magnetic field and the gas must be coupled, so it should 
not be 
expected to occur in regions of low ionization such as the ``dead zone'' 
(Gammie 1996, Turner \& Drake 2009). Therefore, the search for hydrodynamical 
sources of turbulence continues, if only to provide some residual accretion 
in the dead zone. A distinct possibility is the baroclinic instability 
(BI; Klahr \& Bodenheimer 2003, Klahr 2004), the interest in which has been 
recently rekindled (Petersen et al. 2007a, Petersen et al. 2007b, 
Lesur \& Papaloizou 2010). 

A baroclinic flow is one 
where the pressure depends on both density and temperature, 
as opposed to a barotropic flow where the pressure only depends on 
density. In such a flow, the non-axyssimmetric misalignment between 
surfaces of constant density $\rho$ 
(isopycnals) and surfaces of constant pressure $p$ (isobars) 
generates vorticity. Mathematically, 
this translates into a non-zero baroclinic vector, $\curl{(-\rho^{-1}\grad{p})}=
\rho^{-2}\grad{p}\times\grad{\rho}$. Baroclinicity has long been known in atmospheric 
dynamics to be responsible for turbulent patterns on planets 
and for weather patterns of Rossby waves (planetary waves), 
cyclones, and anticyclones on Earth. 

The difference between the baroclinic instability of weather patterns 
on planetary atmospheres and the baroclinic instability in accretion disks 
is that the former is linear, whereas the latter is nonlinear 
(Klahr 2004, Lesur \& Papaloizou 2010). This is because in accretion disks, 
the disturbances have to overcome the strong Keplerian shear that causes 
perturbations to be heavily dominated by restoring forces in all Reynolds numbers.

The nature of the instability was clarified in the work of 
Petersen et al. (2007ab), who highlighted the 
importance of finite thermal inertia. When the thermal time is comparable 
to the eddy turnover time, the vortex is able to establish an entropy gradient 
around itself that compensates the large-scale entropy gradient that 
created it. This entropy gradient back reacts on the eddy, generating 
more vorticity via buoyancy. This in turn reinforces the gradient. A positive feedback 
has been established, and the eddy grows. This, in a nutshell, 
is the baroclinic instability: the sole result of an eddy trying to counter 
the background entropy gradient that established it, and reinforcing 
itself by doing so. 

The 3D properties of the instability have been studied by Lesur \& Papaloizou 
(2010). They find that the vortices produced are not destroyed when 
baroclinicity is present, although they are unstable to the elliptical 
instability (Kerswell 2002, Lesur \& Papaloizou 2009). The saturated state 
of the instability is dominated by 
the presence of large-scale 3D, self-sustained, vortices with weakly 
turbulent cores. The study of Petersen et al. (2007ab) and 
most of that of Lesur \& Papaloizou (2010) was done with spectral codes, 
which filter sound waves. Vortices, however, have the 
interesting property of radiating inertial-acoustic waves, which are known to transport angular 
momentum (Heinemann \& Papaloizou 2009). Lesur \& Papaloizou (2010) performed a compressible, 
yet 2D, simulation, with a resulting Shakura-Sunyaev-like $\alpha$ value of 
$10^{-3}$.

These results are intriguing, and a major question to ask is what their significance is 
in the context of the layered accretion paradigm. Vortices have been described in the 
literature as devoid of radial shear (Klahr \& Bodenheimer 2006), 
so in principle they could form and 
survive in the midst of MRI turbulence, as the simulations of Fromang \& Nelson 
(2005) suggest. Moreover, if the baroclinic instability is 
able to produce and sustain vortices when magnetization is present, synergy with the 
MRI is an interesting possibility, potentially leading to 
higher accretion rates than hitherto achieved in previous works. On the other hand, 
elliptical streamlines can be heavily destabilized by magnetic fields (Lebovitz \& 
Zweibel 2004, Mizerski \& Bajer 2009). This magneto-elliptic instability may either 
be stabilized by baroclinicity, as the elliptic instability was shown to be 
(Lesur \& Papaloizou 2010), or completely break the vortices apart thus rendering 
the baroclinic instability meaningless in the presence of magnetization. We address 
these open questions in this work.

The paper is structured as follows. In \sect{sect:The-model} we introduce the model equations 
of the compressible shearing box, modified to include the contribution from the 
large-scale background entropy gradient. In view of the controversy aroused by the
baroclinic instability in the literature, it was found prudent to establish the 
reliability of the numerics, as well as to provide an independent confirmation 
of the 2D results. This is done in \sect{sect:2D-Results}. The 3D results are 
presented in \sect{sect:3D-Results}. \Tab{table:hydro-disk-models} 
and \Tab{table:magnetic-disk-models} contain summaries of 
the simulations performed for this study, referring to the sections and figures they 
are described. Conclusions are given in \sect{sect:Conclusions}.

\begin{figure*}
  \begin{center}
    \resizebox{.9\textwidth}{!}{\includegraphics{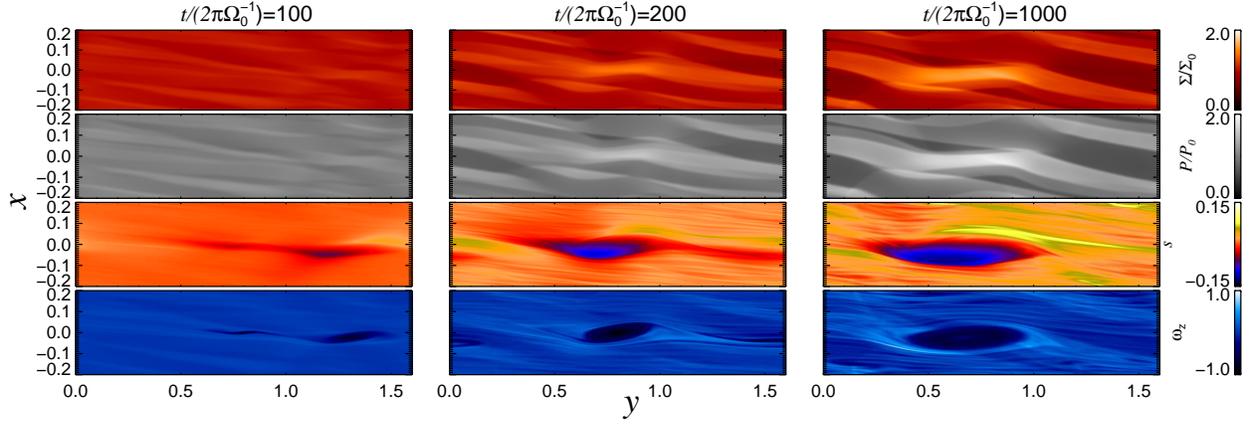}}
\end{center}
\caption[]{Snapshots of the fiducial 2D run with $\xi=2$, $\tau=2\pi$, 
$H=0.1$, and resolution 256$^2$. A vortex is formed and establishes a 
local entropy gradient that counteracts the global entropy gradient that 
caused it in first place. Moderate cooling times keep the surfaces of 
constant density and constant pressure misaligned, leading to more vortex 
growth. In the positive feedback that ensues, giant anti-cyclonic 
vortices grow to the sonic scale. The initial condition was free of 
enstrophy. This vorticity growth was due purely to baroclinic effects.} 
\label{fig:fiducial}
\end{figure*}

\begin{figure*}
  \begin{center}
    \resizebox{.9\hfwidth}{!}{\includegraphics{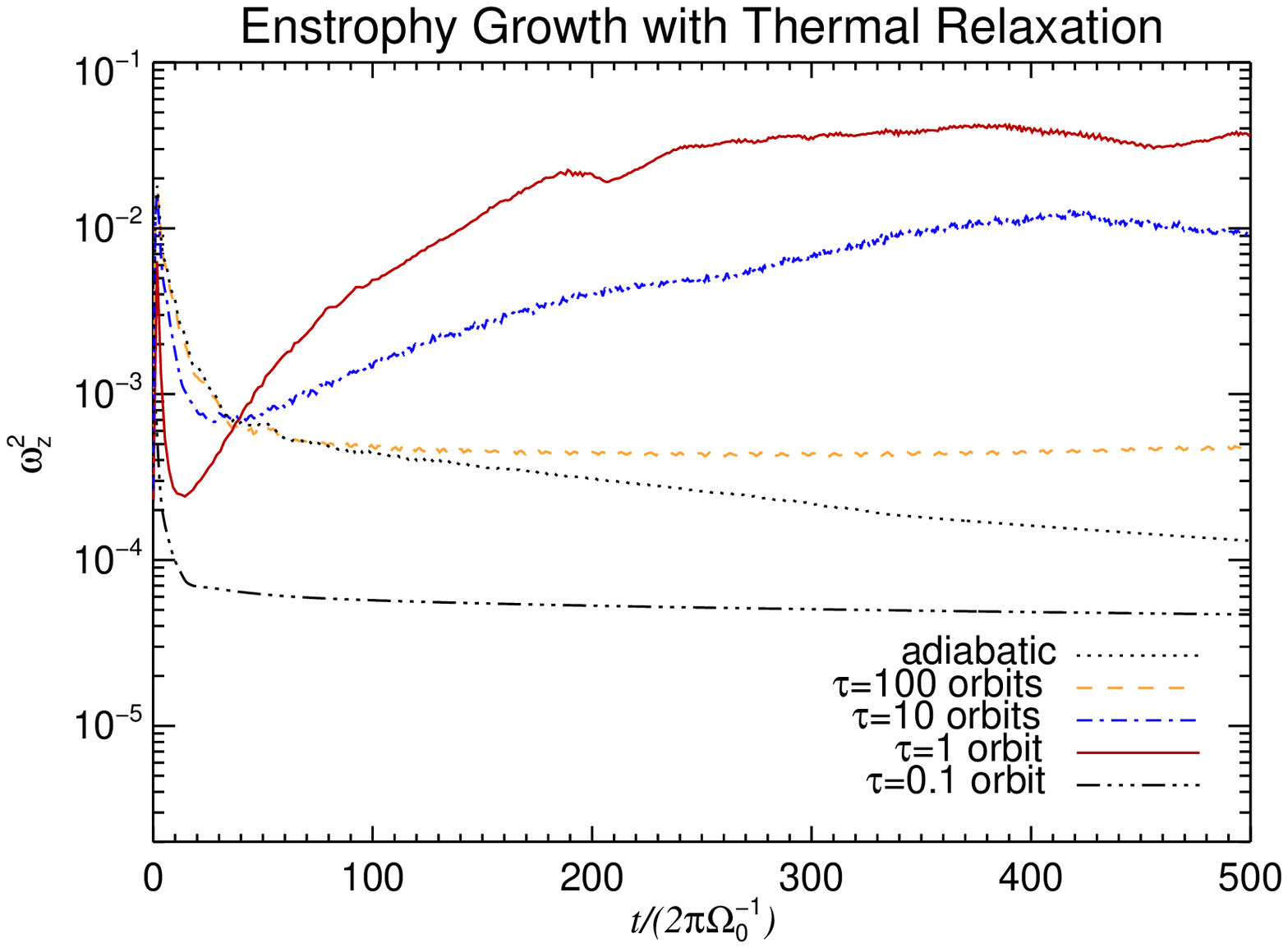}}
    \resizebox{.9\hfwidth}{!}{\includegraphics{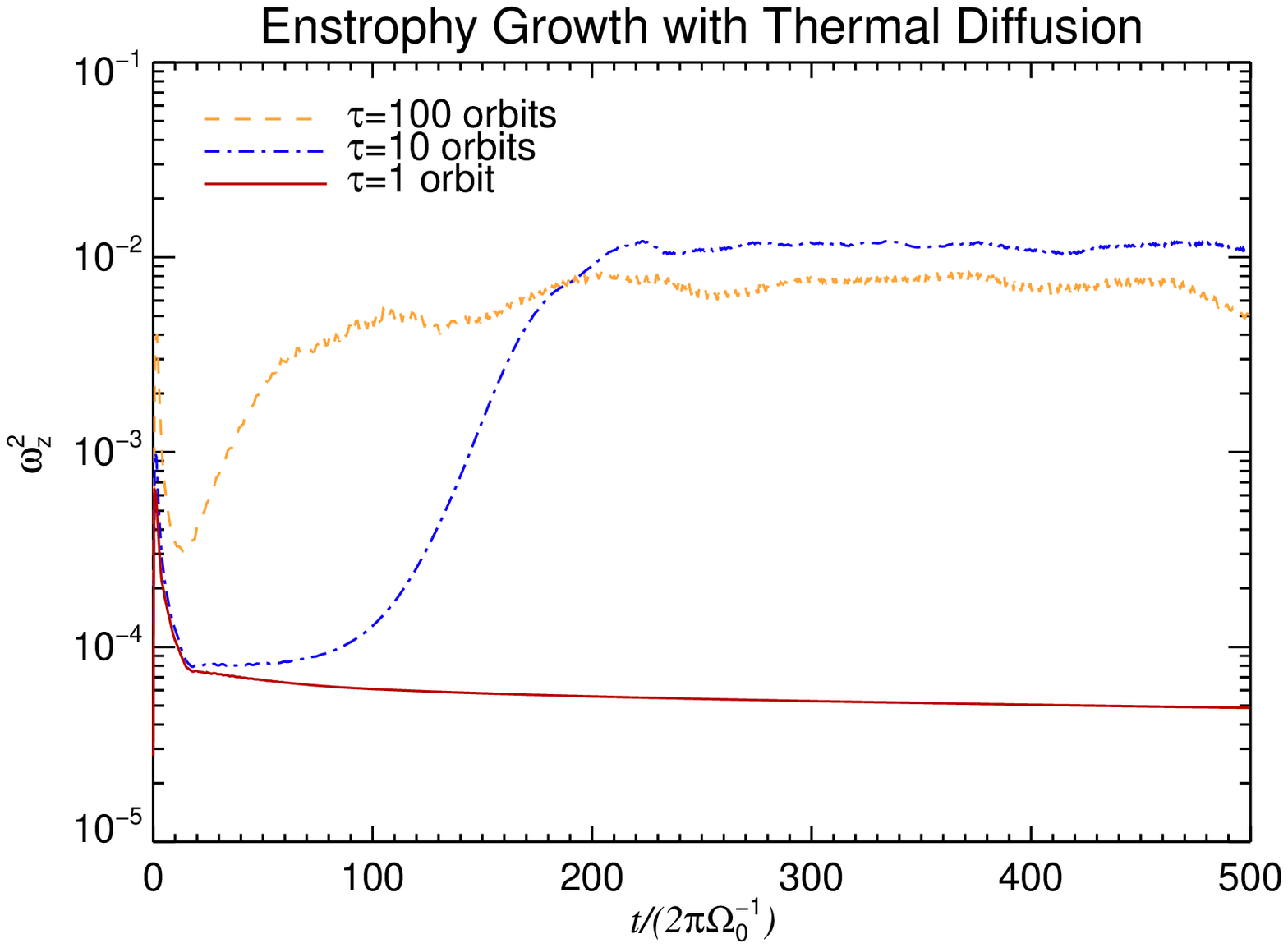}}
\end{center}
\caption[]{{\it Left}). Baroclinic enstrophy growth with thermal relaxation, where a gas parcel 
returns to the initial temperature on a timescale $\tau$. A stable entropy gradient can only be 
maintained between the extremes of too fast a relaxation (isothermal behavior) and too slow 
a relaxation 
(adiabatic behavior). Optimal growth occurs when $\tau$ is comparable to the dynamical time.\\ 
{\it Right}). Baroclinic enstrophy growth with thermal diffusion, where heat diffuses over 
a scale height on a timescale $\tau$. Optimal growth occurs on longer timescales when 
compared to the thermal relaxation case.}
\label{fig:relaxation}
\end{figure*}

\section{The model}
\label{sect:The-model}

We model a local patch of the disk following the shearing box approximation.  
The reader is referred to Regev \& Umurhan (2008) for an extensive discussion 
of the limitations of the approximation. To include the baroclinic term, we 
consider a large-scale radial pressure gradient following a power law of index $\xi$
 
\[\bar{p}(r) = p_0 (r/R_0)^{-\xi} ,\]where $r$ is the cylindrical radius and $R_0$ 
is a reference radius. The overbar indicates that this quantity is time-independent.
The total pressure is $p_{\rm tot}$=$\bar{p}+p$, where $p$ is the local fluctuation.
The linearization of this gradient is done in the same way as the 
large-scale Keplerian shear is linearized in the shearing box. This 
leads to extra terms in the equations involving the radial pressure 
gradient. We quote the modified shearing box equations below. A 
derivation of the extra terms is presented in \app{app:linearization}.

\begin{eqnarray}
%
%
  \aderiv{\rho} &+&\left(\v{u}\cdot\del\right)\rho = -\rho\Div\v{u} + f_D(\rho)\label{eq:continuity}\\
%
%
  \aderiv{\v{u}}&+&\left(\v{u}\cdot\del\right)\v{u} =-\frac{1}{\rho}\grad{p} + \frac{\v{J}\times\v{B}}{\rho} -2\varOmega_0\left(\hatz\times\v{u}\right)\nonumber\\
&&+ \frac{3}{2}\varOmega_0u_x\haty + \frac{\xi p_0}{R_0}\left(\frac{1}{\rho} - \frac{1}{\rho_0}\right)\hatx + f_\nu(\v{u},\rho) \label{eq:navier-stokes}\\
%
%
 \aderiv{\v{A}} &=& \frac{3}{2}\varOmega_0A_y\hatx + \v{u}\times\v{B} -\mu_0 \eta \v{J} + f_\eta(\v{A})\label{eq:induction}\\
%
%
 \aderiv{s}&+&\left(\v{u}\cdot\del\right)s =\frac{1}{\rho T}\left\{\Div\left(K\grad{T}\right) - \rho \cv\frac{\left(T-T_0\right)}{\tau_c} \white{\frac{1}{1}}\right.\nonumber\label{eq:entropyeq}\\
&&+\left. \mu_0\eta\v{J}^2 + \frac{\xi p_0}{R_0}\frac{u_x}{(\gamma-1)} \right\} + f_K(s).
\end{eqnarray} 

In the equations above, $\v{u}$ is the velocity, $\v{A}$ the magnetic vector potential,  $\v{B}$=$\curl{\v{A}}$ 
the magnetic field, $\v{J}$=$\curl{\v{B}}/\mu_0$ the current (where $\mu_0$ is the magnetic 
permittivity), $\eta$ is the resistivity, $T$ the temperature, $s$ the entropy, and $K$ the 
radiative conductivity. The operator 

\begin{equation}
  \aderiv{}=\pderiv{}{t} + u_y^{(0)}\pderiv{}{y}
\end{equation}represents the Keplerian derivative of a fluid parcel. It is the only place where the Keplerian flow $u_y^{(0)}$ appears explicitly. The 
advection is made Galilean-invariant by means of the SAFI algorithm 
(Johansen et al. 2009), which speeds up performance. The simulations 
were done with the {\sc Pencil Code} 
{\footnote{See http://www.nordita.org/software/pencil-code/}}. 

We work with entropy as the main thermodynamical quantity. This is a natural choice when 
dealing with baroclinicity. Considering the polytropic equation of state, 

\begin{equation}
  p=k\rho^{(n+1)/n},
\end{equation}and the definition of entropy,

\begin{equation}
  s=\cv\left[\ln(p/p_0)-\gamma\ln(\rho/\rho_0)\right],
  \label{eq:entropy}
\end{equation}we immediately recognize $s$=$\cv\ln(k/k_0)$ in the 
case $n$=$1/(\gamma-1)$; i.e., up to a constant, entropy is the proportionality
factor in the polytropic equation of state.  That means that any spatial 
gradient of entropy translates into a departure from barotropic conditions 
{\footnote{Actually, this 
is such a useful insight that some authors prefer to define entropy as
$S$=$p/\rho^\gamma$. The reader should then keep in mind that what we call 
entropy is actually $s$=$\cv\ln(S/S_0)$ in that definition.  Here we prefer 
to use the definition \eq{eq:entropy} as it comes from thermodynamics;
i.e., $Tds$=$de+pdv$, where $e$=$\cv T$ is the internal energy and 
$v$=$1/\rho$. It also enables the 
Brunt-V\"ais\"al\"a frequency to be written in a more compact 
form (\eq{eq:bv}).}}.

The third term on the right-hand side of the entropy equation is an artificial 
thermal relaxation term, which drives the temperature back to the initial temperature $T_0$,
on a pre-specified thermal timescale $\tau_c$. The temperature is 
$T$=$c_s^2/\left[c_p(\gamma-1)\right]$, where $c_s$ is the sound speed, 
$\gamma$=$c_p/\cv$ the adiabatic index, and $\cv$ and $c_p$ are the 
heat capacities at constant volume and constant pressure, respectively. 
 
After defining entropy, we also define the Brunt-V\"ais\"al\"a frequency $N$, 
the frequency associated with buoyant structures

\begin{eqnarray}
  N^2&=&-\frac{1}{c_p\rho}\grad{p}\cdot\grad{s}\label{eq:bv}\\
     &=&\frac{1}{\rho}\pderiv{p}{r}\left(\frac{1}{\rho}\pderiv{\rho}{r} -
       \frac{1}{\gamma p}\pderiv{p}{r}\right),\nonumber
\end{eqnarray}and we have assumed axis-symmetry ($\partial_\phi$=0) 
and no vertical stratification ($\partial_z$=0) between the steps. 
In our setup, there is no large-scale density 
gradient, so the first term inside the parentheses cancel. As 
$dp/dr$=-$p\ksi/r$, we have, at $r$=$R_0$

\begin{equation}
  N^2 = -\left(\frac{c_{s0}\,\ksi}{\gamma R_0}\right)^2;
\end{equation}i.e., the Brunt-V\"ais\"al\"a frequency is always imaginary. 
However, the flow is convectively stable, since the epicyclic frequency squared 

\begin{equation}
  \kappa^2 = \frac{1}{r^3}\frac{d j^2}{dr}
  \label{eq:kappa}
\end{equation}is far higher than $-N^2$, so that the Solberg-H{\o}iland 
criterion is always satisfied

\begin{equation}
  \kappa^2 + N^2 > 0.
\end{equation} In \eq{eq:kappa}, $j$=$\varOmega r^2$ is the 
specific angular momentum per unit mass. 

We add explicit sixth-order hyperdiffusion $f_D(\rho)$, hyperviscosity $f_\nu(\v{u},\rho)$, 
and hyperresistivity $f_\eta(\v{A})$ to the mass, momentum, and induction equations as specified in 
Lyra et al (2008). Hyper heat conductivity $f_K(s)$ to the entropy equation is added as 
in Lyra et al. (2009) and Oishi \& Mac Low (2009). All simulations use 
$c_p$=$R_0$=$\varOmega_0$=$\rho_0$=$\mu_0$=1, $\gamma$=1.4, and $c_s$=0.1. 

\section{2D Results}
\label{sect:2D-Results}

\begin{figure}
  \begin{center}
    \resizebox{.9\hfwidth}{!}{\includegraphics{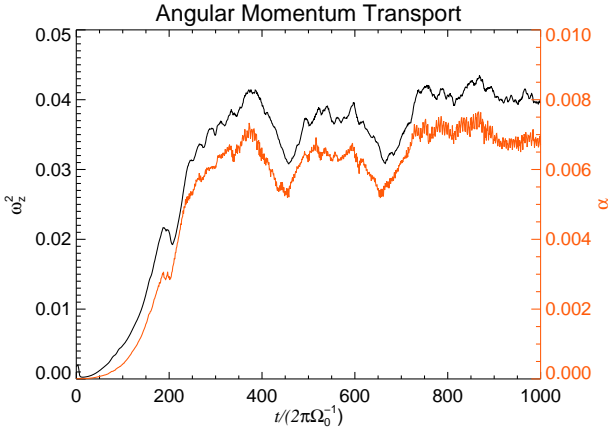}}
\end{center}
\caption[]{Enstrophy and the resulting alpha-viscosity for the fiducial 2D run. The two 
quantities are quite well correlated, since the angular momentum transport is the 
result of inertial-acoustic waves, which in turn are driven by vorticity.} 
\label{fig:alpha-enstrophy}
\end{figure}


\begin{figure}
  \begin{center}
    \resizebox{.9\hfwidth}{!}{\includegraphics{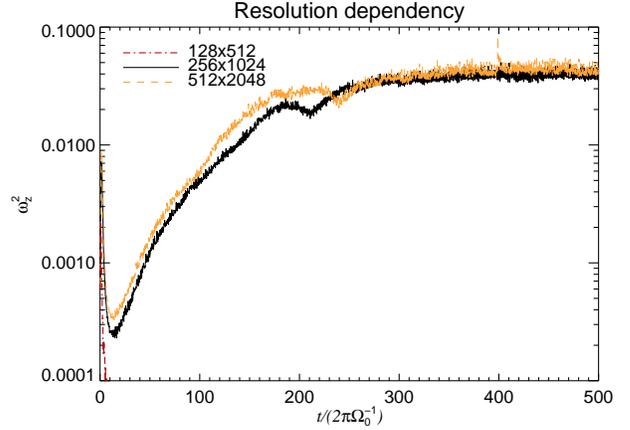}}
\end{center}
\caption[]{Dependence on resolution. The low resolution run 
fails to develop vortices, reaffirming that aliasing is not 
occurring in our models. The middle and high resolution runs saturate 
at the same level of enstrophy, which suggests convergence.}
\label{fig:resolution}
\end{figure}

\begin{figure}
  \begin{center}
    \resizebox{.9\hfwidth}{!}{\includegraphics{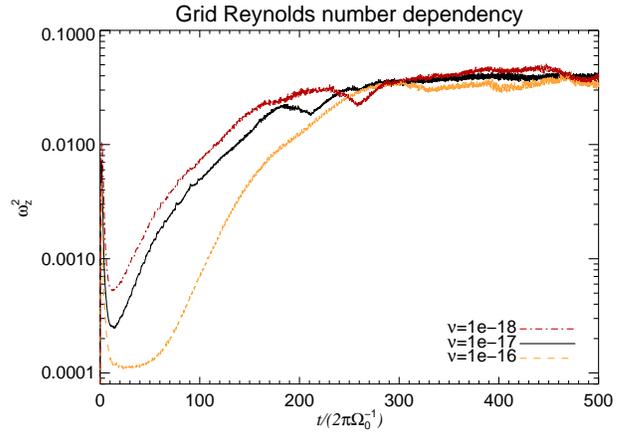}}
\end{center}
\caption[]{Dependence on Grid Reynolds number. The hyperviscosities 
shown correspond to Re=0.002, 0.02, and 0.2 with respect to the 
velocity shear introduced by the Keplerian flow, calculated on 
the grid scale. The initial phase of growth occurs at Re$<$1, where 
it is seen that the amount of growth depends on the Reynolds number. 
Upon saturation, all simulations converge to the same level of enstrophy. 
A heavily aliased solution occurs for $\nu^{(3)}$=$\ttimes{-21}$, where even 
a simulation seeded only with noise develops vortices. The same does 
not occur for the hyperviscosities shown, where finite amplitude 
perturbations were required. We usually use 
$\nu^{(3)}$=$\ttimes{-17}$.}
\label{fig:reynolds}
\end{figure}

\begin{table*}
\caption[]{Simulation suite parameters for nonmagnetic runs and results.}
\begin{center}
\begin{tabular}{lcc ccc c ccc cc c} \hline \hline
&\multicolumn{5}{c} {\sc Parameter} & & \multicolumn{6}{c}{\sc Results}\\\cline{2-6}\cline{8-13}
\multirow{2}{*}\tiny{\sc Run} &\tiny $\tau_c$&\tiny $k_{\rm th}$ &\tiny $\ksi$ &\tiny Resolution &\tiny $\nu^{(3)}$ &\tiny&\tiny $<$$\omega_z^2$$>$&\tiny$\alpha$&\tiny $<$$u^2$$>$&$\rho_{\rm rms}$ &\tiny $<$$T$$>$&\tiny$<$$u_z^2$$>$ \\
&\tiny($/t_{\rm orb}$)&&&&&&\tiny($/\varOmega_K^2$)&&\tiny($/c_{s0}^2$)&\tiny($/\rho_0$)&\tiny($/T_0$)&\tiny($/c_{s0}^2$)\\\hline
%
\multicolumn{13}{c}{\multirow{2}{*}{Fiducial 2D , \sect{sect:Baroclinic-production-of-vorticity}, \Figs{fig:fiducial}{fig:alpha-enstrophy}}}\\\\\hline
\tiny A &\tiny   1    &\tiny0  &\tiny 2&\tiny256$^2$    &\tiny$\ttimes{-17}$&\tiny&
          \tiny 0.04&\tiny 0.007&\tiny 0.05&\tiny 0.03&\tiny 1.01&\tiny-- \\\hline
\multicolumn{13}{c}{\multirow{2}{*}{Thermal relaxation, \sect{sect:Thermal-time}, \fig{fig:relaxation}}}\\\\\hline
\tiny B &\tiny  0.1   &\tiny0  &\tiny 2&\tiny256$^2$    &\tiny$\ttimes{-17}$&\tiny&
          \tiny$\xtimes{5}{-5}$&\tiny 0&\tiny $\xtimes{2}{-6}$&\tiny 0.001&\tiny1 &\tiny --\\
\tiny C &\tiny  10    &\tiny0  &\tiny 2&\tiny256$^2$    &\tiny$\ttimes{-17}$&\tiny&
          \tiny 0.01&\tiny 0.001&\tiny 0.01&\tiny 0.01&\tiny 1.01&\tiny--\\
\tiny D &\tiny 100    &\tiny0  &\tiny 2&\tiny256$^2$    &\tiny$\ttimes{-17}$&\tiny&
          \tiny$\xtimes{6}{-4}$ &\tiny $\ttimes{-5}$&\tiny $\xtimes{2}{-4}$&\tiny 0&\tiny0.99 &\tiny --\\
\tiny E &\tiny$\infty$&\tiny0  &\tiny 2&\tiny256$^2$    &\tiny$\ttimes{-17}$&\tiny&
          \tiny $\xtimes{2}{-4}$&\tiny 0&\tiny $\xtimes{2}{-5}$&\tiny 0.001&\tiny 0.98&\tiny -- \\\hline
\multicolumn{13}{c}{\multirow{2}{*}{Thermal diffusion, \sect{sect:Diffusion}, \fig{fig:relaxation}}}\\\\\hline
\tiny F &\tiny$\infty$&\tiny1  &\tiny 2&\tiny256$^2$    &\tiny$\ttimes{-17}$&\tiny&
          \tiny$\xtimes{5}{-5}$ &\tiny 0 &\tiny $\xtimes{2}{-6}$&\tiny 0.001&\tiny 1.03&\tiny --\\
\tiny G &\tiny$\infty$&\tiny10 &\tiny 2&\tiny256$^2$    &\tiny$\ttimes{-17}$&\tiny&
          \tiny 0.01&\tiny 0.004&\tiny 0.03&\tiny 0.001&\tiny 12&\tiny--\\
\tiny H &\tiny$\infty$&\tiny100&\tiny 2&\tiny256$^2$    &\tiny$\ttimes{-17}$&\tiny&
          \tiny 0.007&\tiny 0.001&\tiny 0.01&\tiny 0.003&\tiny 3&\tiny--\\\hline
\multicolumn{13}{c}{\multirow{2}{*}{Resolution, \sect{sect:Resolution}, \fig{fig:resolution}}}\\\\\hline
\tiny J &\tiny   1    &\tiny0  &\tiny 2&\tiny128x512    &\tiny$\xtimes{3}{-15}$&\tiny&
          \tiny $\xtimes{7}{-6}$ &\tiny 0&\tiny $\xtimes{2}{-6}$&\tiny 0&\tiny 1&\tiny --\\
\tiny K &\tiny   1    &\tiny0  &\tiny 2&\tiny256x1024   &\tiny$\ttimes{-17}$&\tiny&
          \tiny 0.04&\tiny 0.007&\tiny 0.05&\tiny 0.05&\tiny 1.01&\tiny --\\
\tiny L &\tiny   1    &\tiny0  &\tiny 2&\tiny512x2048   &\tiny$\xtimes{3}{-18}$&\tiny&
          \tiny 0.04&\tiny 0.008&\tiny 0.06&\tiny 0.05&\tiny 1 &\tiny -- \\\hline
\multicolumn{13}{c}{\multirow{2}{*}{Reynolds number, \sect{sect:Grid-Reynolds-number}, \fig{fig:reynolds}}}\\\\\hline
\tiny M &\tiny   1    &\tiny0  &\tiny 2&\tiny 256x1024   &\tiny$\ttimes{-16}$&\tiny&
          \tiny 0.03&\tiny 0.007&\tiny 0.05&\tiny 0.04&\tiny 1&\tiny --\\
\tiny N &\tiny   1    &\tiny0  &\tiny 2&\tiny256x1024   &\tiny$\ttimes{-18}$&\tiny&
          \tiny 0.04&\tiny 0.006&\tiny 0.05&\tiny 0.04&\tiny 1.01&\tiny --\\\hline
\multicolumn{13}{c}{\multirow{2}{*}{Fiducial 3D , \sect{sect:3D-Results}, \Figssss{fig:evolution-oo-ss}{fig:3d-timeseries}{fig:3d-vertical-slice}{fig:vortex-shear}{fig:magruns}}}\\\\\hline
\tiny O &\tiny   1    &\tiny0  &\tiny 2&\tiny256$^2$x128&\tiny$\ttimes{-17}$&\tiny&
          \tiny 0.03&\tiny 0.004&\tiny 0.03&\tiny 0.02&\tiny 1.01&\tiny$\xtimes{3}{-5}$ \\\hline
\end{tabular}
\end{center}
\label{table:hydro-disk-models}
\end{table*}

\begin{table}
\caption[]{Simulation suite parameters for magnetic runs.}
\begin{center}
\begin{tabular}{lcc c cc} \hline \hline
\multirow{2}{*}\tiny{\sc Run} &\tiny $\tau_c$&\tiny $k_{\rm th}$ &\tiny $\ksi$ & \tiny $B_0$ & \tiny $\eta$ \\\hline

\multicolumn{6}{c}{\multirow{2}{*}{MRI, \sect{sect:Magnetic-fields}, \Figsss{fig:evolution-mribi-3b}{fig:evolution-mribi-boxaverage}{fig:evolution-oo-bb}{fig:magruns}}}\\\\\hline
\tiny P &\tiny   1    &\tiny0  &\tiny 2&\tiny$\xtimes{5}{-3}$&\tiny0 \\\hline
\multicolumn{6}{c}{\multirow{2}{*}{Strong field, \sect{sect:Increasing-the-field-strength}, \Figs{fig:magruns}{fig:channel-flow}}}\\\\\hline
\tiny Q1 &\tiny   1    &\tiny0  &\tiny 2&\tiny$\xtimes{6}{-2}$&0 \\\hline
\tiny Q2 &\tiny   1    &\tiny0  &\tiny 2&\tiny$\xtimes{3.75}{-2}$&\tiny$\xtimes{8.8}{-4}$ \\\hline
\multicolumn{6}{c}{\multirow{2}{*}{Weak field, \sect{sect:Decreasing-the-field-strength}, \Figs{fig:magruns}{fig:vortex-mri}}}\\\\\hline
\tiny R &\tiny   1    &\tiny0  &\tiny 2&\tiny$\xtimes{1.5}{-3}$&\tiny0 \\\hline
\multicolumn{6}{c}{\multirow{2}{*}{Resistivity, \sect{sect:Resistivity}, \Figss{fig:magruns}{fig:resistivity}{fig:resistivity-xyplane}}}\\\\\hline
\tiny S1 &\tiny   1    &\tiny0  &\tiny 2&\tiny$\xtimes{5}{-3}$&\tiny$\ttimes{-3}$ \\\hline
\tiny S2 &\tiny   1    &\tiny0  &\tiny 2&\tiny$\xtimes{5}{-3}$&\tiny$\xtimes{1.6}{-4}$ \\\hline
\multicolumn{6}{c}{\multirow{2}{*}{Azimuthal field, \sect{sect:other-fields}, \fig{fig:spacetime}}}\\\\\hline
\tiny T &\tiny   1    &\tiny0  &\tiny 2&\tiny$\xtimes{3}{-2}$&\tiny0 \\\hline
\multicolumn{6}{c}{\multirow{2}{*}{Zero net flux field, \sect{sect:other-fields}, \fig{fig:spacetime}}}\\\\\hline
\tiny U &\tiny   1    &\tiny0  &\tiny 2&\tiny$\times{-2}$&\tiny0 \\\hline
\multicolumn{6}{c}{\multirow{2}{*}{Control runs, \sect{sect:Magnetic-fields}, \fig{fig:spacetime}}}\\\\\hline
\tiny CP &\tiny   1    &\tiny0  &\tiny 0&\tiny$\xtimes{5}{-3}$&\tiny0 \\\hline
\tiny CR &\tiny   1    &\tiny0  &\tiny 0&\tiny$\xtimes{1.5}{-3}$&\tiny0 \\\hline
\tiny CS &\tiny   1    &\tiny0  &\tiny 0&\tiny$\xtimes{5}{-3}$&\tiny$\xtimes{1.6}{-4}$ \\\hline
\hline
\end{tabular}
\end{center}
\label{table:magnetic-disk-models}
\end{table}

Although the most interesting results for the effects of the 
baroclinic generation of vorticity should be given by 3D models, the 
existence of the baroclinic instability has been strongly contested even in 
two dimensions. We therefore consider it important to present 2D results 
confirming its excitation. In \fig{fig:fiducial} we present a fiducial 2D
run where the evolution of the baroclinic instability is followed. It 
corresponds to run A in \Tab{table:hydro-disk-models}. The slope of the entropy 
gradient is $\ksi=2$, which corresponds to a Richardson number 
$\Ri= 4N^2/9\varOmega_K^2 \simeq -\xtimes{9}{-3}$. As initial condition, 
we seed the box with noise at small wavenumbers only, following 

\begin{eqnarray}
\varSigma(x,y)&=&C\varSigma_0{\rm e}^{-(x/2\sigma)^2} \times \nonumber\\
&&\sum_{i=-k_x}^{k_x}\sum_{j=0}^{k_y}{\sin\left\{2\pi\left(i\frac{x}{L_x} + j\frac{y}{L_y} + \phi_{ij}\right)\right\}}.
\label{eq:initcond}
\end{eqnarray}The phase $0<\phi<1$ determines the randomness. The subscripts 
underscore that the phase is the same for all grid points, only changing 
with wavenumber. The constant $C$ sets the strength of the perturbation. As 
stressed by Lesur \& Papaloizou (2010), the baroclinic instability is 
subcritical, and therefore a finite initial amplitude is needed 
to trigger growth. We set the constant $C$ to 
yield $\varSigma_{\rm rms}=0.05$. The entropy is then initialized such 
that $p$=$p_0$$\equiv$$\cte$ in the sheet. 

The rationale behind this unorthodox initial condition is that this noise 
is independent of resolution. The usual Gaussian noise distributes power 
through all wavelengths, so the wavelengths from $k<10$ are assigned increasingly less power as the 
resolution increases. We stress that it is not vital for the instability 
to be seeded with resolution-independent noise, nor are we missing important 
physics by not exciting the small scales. 

We do not seed noise in the velocity field. The initial 
condition is strictly nonvortical. Since in 2D the stretching term 
is absent, any increase in vorticity can only be a baroclinic effect. 

\subsection{Baroclinic production of vorticity}
\label{sect:Baroclinic-production-of-vorticity}

The baroclinic term, in 2D, is 

\begin{equation}
  \left(\grad{p}\times\grad\rho\right)_z = \partial_xp\partial_y\rho - \partial_x\rho\partial_y{p} - \xi{p_0}R_0^{-1}\partial_y\rho .
\end{equation}The two first terms are local, whereas the third comes from the 
large-scale gradient. There would be a fourth if we considered a large-scale 
density gradient as well. This third term generates vorticity out of any azimuthal 
perturbation in the density, much in the same way as the locally isothermal approximation 
does in global disks. This term is paramount, since it is the only source 
term that will generate net enstrophy in the flow. In the beginning, this term 
dominates, generating enstrophy out of the initial density perturbations. The 
enstrophy is then amplified by the local baroclinic vector via the positive 
feedback described in the introduction.

We witness the same general phenomena as Petersen et al. (2007a), even 
though the details of implementating the entropy gradient are different. 
In global simulations, the vortex swings gas parcels back 
and forth from cold to hot, which causes baroclinicity and vortex growth.
Here the initial temperature 
all over the box is the same, so the vortex does not automatically swing gas from cold 
to hot. It swings it up and down the hard-coded pressure(=temperature=entropy) 
gradient. The second-to-last term on the right-hand side of 
the entropy equation comes from the linearization of the advection term in 
the presence of an entropy gradient. It embodies how the relative entropy 
of a fluid parcel with respect to the background entropy changes as it walks 
up or down an entropy gradient, clearly 
demonstrated by the dependence on $u_x$. Because the movement along the 
vortex lines is embedded with a $u_x$ component of motion, this term 
increases or decreases the entropy within the gas parcel depending on the sign of $u_x$. 
Of course, this is just the same physical effect in a different frame of reference. 

We show a time series of the flow in \fig{fig:fiducial} as snapshots 
of density, pressure, entropy, and vorticity. From these snapshots 
we see that the density and 
the pressure are very correlated. One would therefore expect that the 
amount of baroclinicity produced is tiny or vanishingly small. However, 
looking at the snapshots of entropy, we see that 
appearances are deceiving: the vortex generates a strong radial entropy 
gradient around itself. This is what we described qualitatively in the 
introduction, and what Petersen et al. (2007ab) called a ``sandwich 
pattern''. Notice that the sign is indeed reversed with respect to 
the global gradient (higher at negative values of $x$). This pattern 
of a local entropy gradient developed by the vortex is a constant 
feature throughout the simulations.

\subsection{Angular momentum transport}
\label{sect:Angular-momentum-transport}

A very important question to ask is what is the strength of the 
angular momentum transport of the resulting baroclinic unstable flow. 
We measured the kinetic alpha values in the simulations, 

\begin{equation}
  \alpha=\frac{R^{xy}}{qp_0}
\end{equation}where $R^{xy}$=$\overline{\rho\delta{u_x}\delta{u_y}}$ is the 
Reynolds stress. The value measured is $\alpha\approx\xtimes{5}{-3}$, 
indicating good transport of angular 
momentum. The temporal variation of alpha (\fig{fig:alpha-enstrophy}) 
matches that of the enstrophy. 

This correlation is understood in light of the shear-vortex wave coupling. The 
angular momentum transport does not come from the vortex itself, but is instead caused 
by the inertial-acoustic waves that are driven by vorticity. For a detailed 
explanation, see Mamatsashvili \& Chagelishvili (2007), Heinemann \& Papaloizou (2009), 
and Tevzadze et al. (2010). The same production of shear waves and associated angular 
momentum transport are seen in the 2D compressible runs of Lesur \& Papaloizou (2010). 
It should be kept in mind that the quoted angular momentum transport may 
be overestimated because of using a shearing box. As pointed out 
by Regev \& Umurhan (2008), the shearing box approximation may 
lead to wrong results because it has a limited spatial 
scale, excessive symmetries, and uses periodic boundaries. 
In the particular case of a vortex in a box, the periodic boundaries 
enforce interaction between the vortex and the strain field of its 
own images, which may lead to spurious generation of Reynolds stress.

We underscore again that the initial condition was nonvortical. The finite-amplitude
perturbations are turned into vortical patches by the global baroclinic term, which then 
may more owing to the local baroclinic feedback. 

Other point worth highlighting is that the instability has slow growth rates on 
the order of a hundred orbits. The saturated state is only weakly compressible, 
with the rms density $<\rho^2> - <\rho>^2$ at a modest 0.05. 

All our simulations are 
compressible, so they are timestep limited by the presence of sound waves. The viscosity 
and heat diffusion are explicit, so they influence the Courant condition, which further 
limits the timestep. We calculated the fiducial model for 1000 orbits, 
but the other runs, which intend to explore the parameter space, were only calculated 
for 500. These are shown in the next sections. 

\subsection{Thermal time}
\label{sect:Thermal-time}

The fiducial simulation had a constant thermal relaxation time equal to 
one orbit, $\tau_c$=$\tau_{\rm orb}$ in \eq{eq:entropy}, where 
$\tau_{\rm orb}$=$2\pi/\varOmega$. This is more representative 
of the very outer disk, $\approx$100\,AU,  but at 10\,AU the disk is optically thick. 
The thermal time is therefore expected to be much longer, and it is 
instructive to examine the behavior of the baroclinic instability in such a regime. 
We ran simulations with thermal times 
of 10 and 100 orbits, shown in \fig{fig:relaxation}. These correspond to runs C and D 
in \Tab{table:hydro-disk-models}. The extreme cases of an adiabatic 
run (run E) and a nearly isothermal one ($\tau_c$=0.1 orbit, run B) are also presented. 
In agreement with Petersen et al. (2007a), the runs with longer thermal times allow 
for a stronger increase in enstrophy in the first orbits, also seen in the 
adiabatic case. This is because the initial thermal 
perturbations disperse slowly without thermal relaxation, thus remaining tight 
(strong gradients) and allowing for a stronger baroclinic amplification. 

After the first eddies appear, the establishment of a baroclinic feedback needs a 
fast cooling time to lead to the reverted entropy gradient seen in the fiducial run. The 
most vigorous enstrophy growth in this phase is indeed seen to be the one with $\tau_c$ equal 
to one orbit. For a cooling time of 10 orbits, sustained growth of enstrophy only happens at 
later times (between 20 and 50 orbits, as opposed to 10 orbits for $\tau_c$=$\tau_{\rm orb}$),  
and leads to 5 times less (grid-averaged) enstrophy at 150 orbits. The isothermal case and the adiabatic 
cases, as expected, can never establish the counter entropy gradient needed for the baroclinic 
feedback and do not experience enstrophy growth past the initial phase. 

\begin{figure*}
  \begin{center}
    \resizebox{.285\textwidth}{!}{\includegraphics{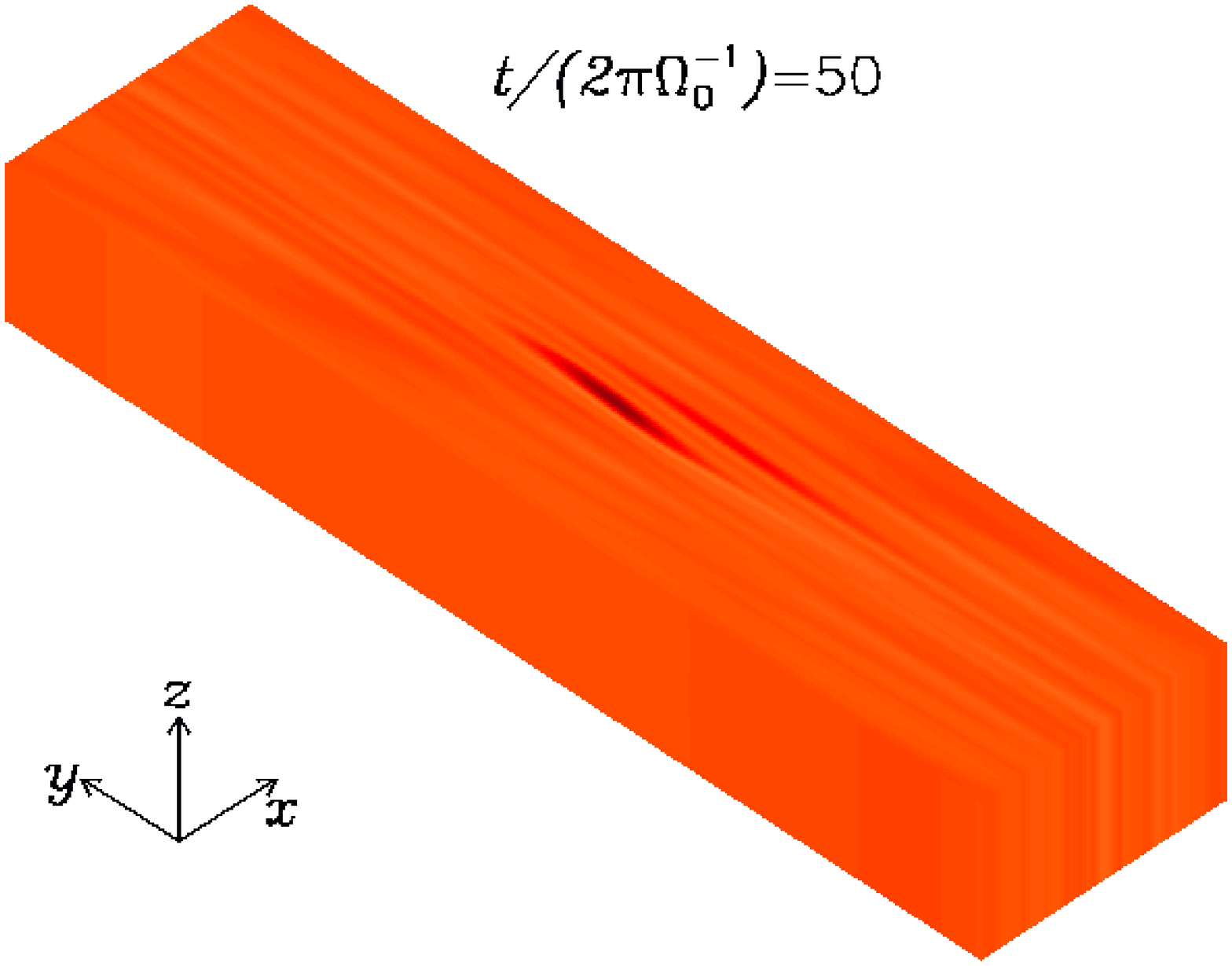}}
    \resizebox{.285\textwidth}{!}{\includegraphics{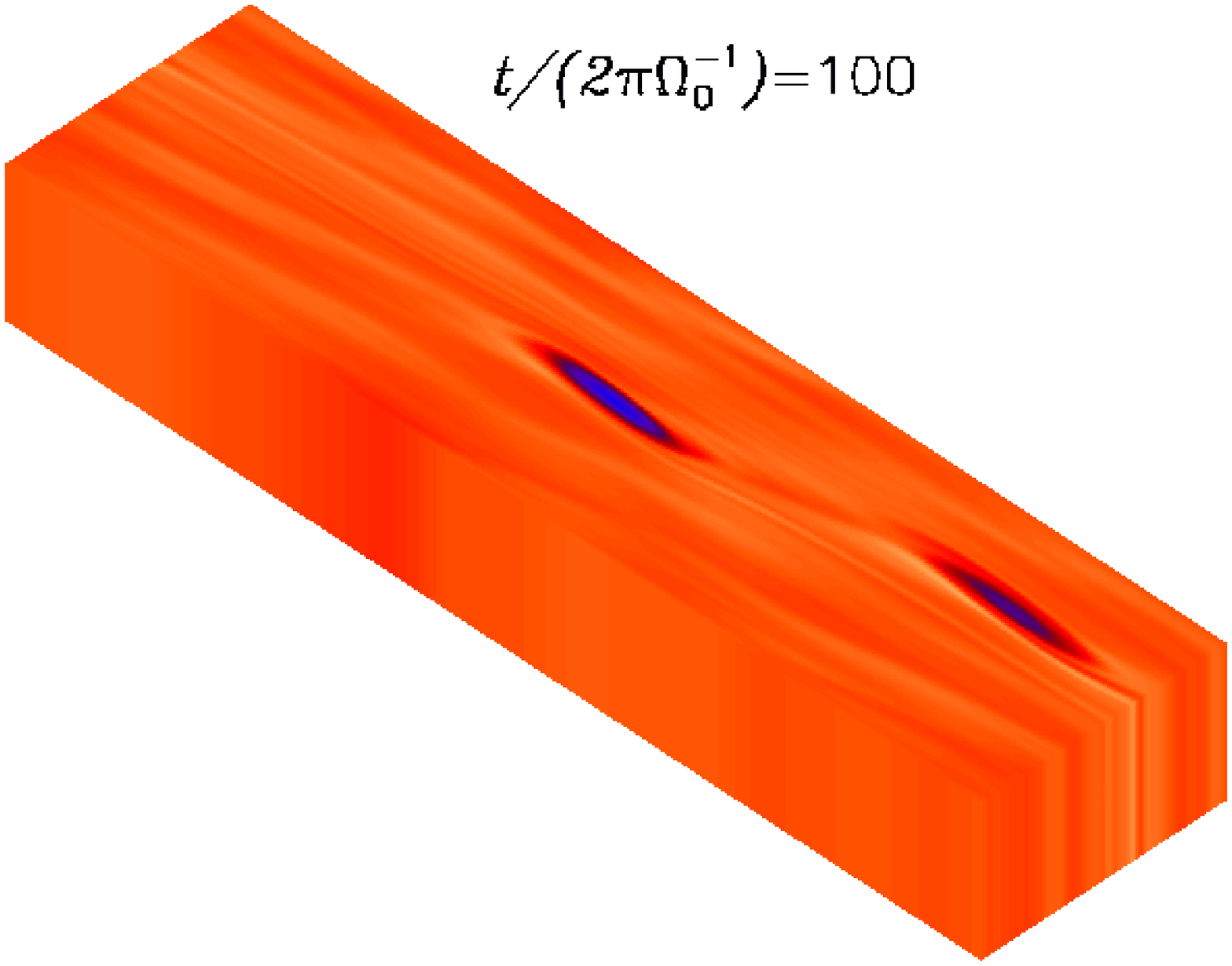}}
    \resizebox{.285\textwidth}{!}{\includegraphics{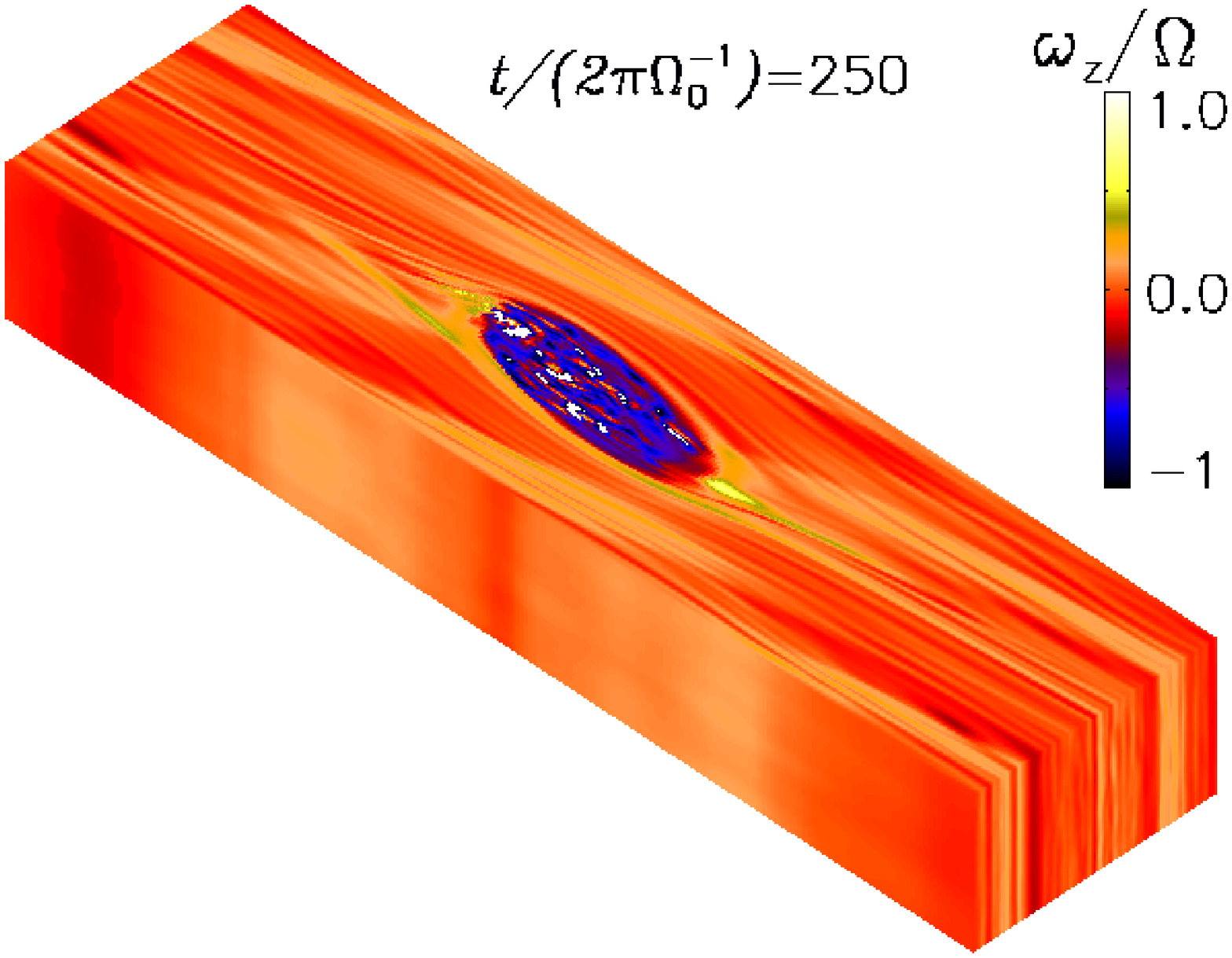}}
    \resizebox{.285\textwidth}{!}{\includegraphics{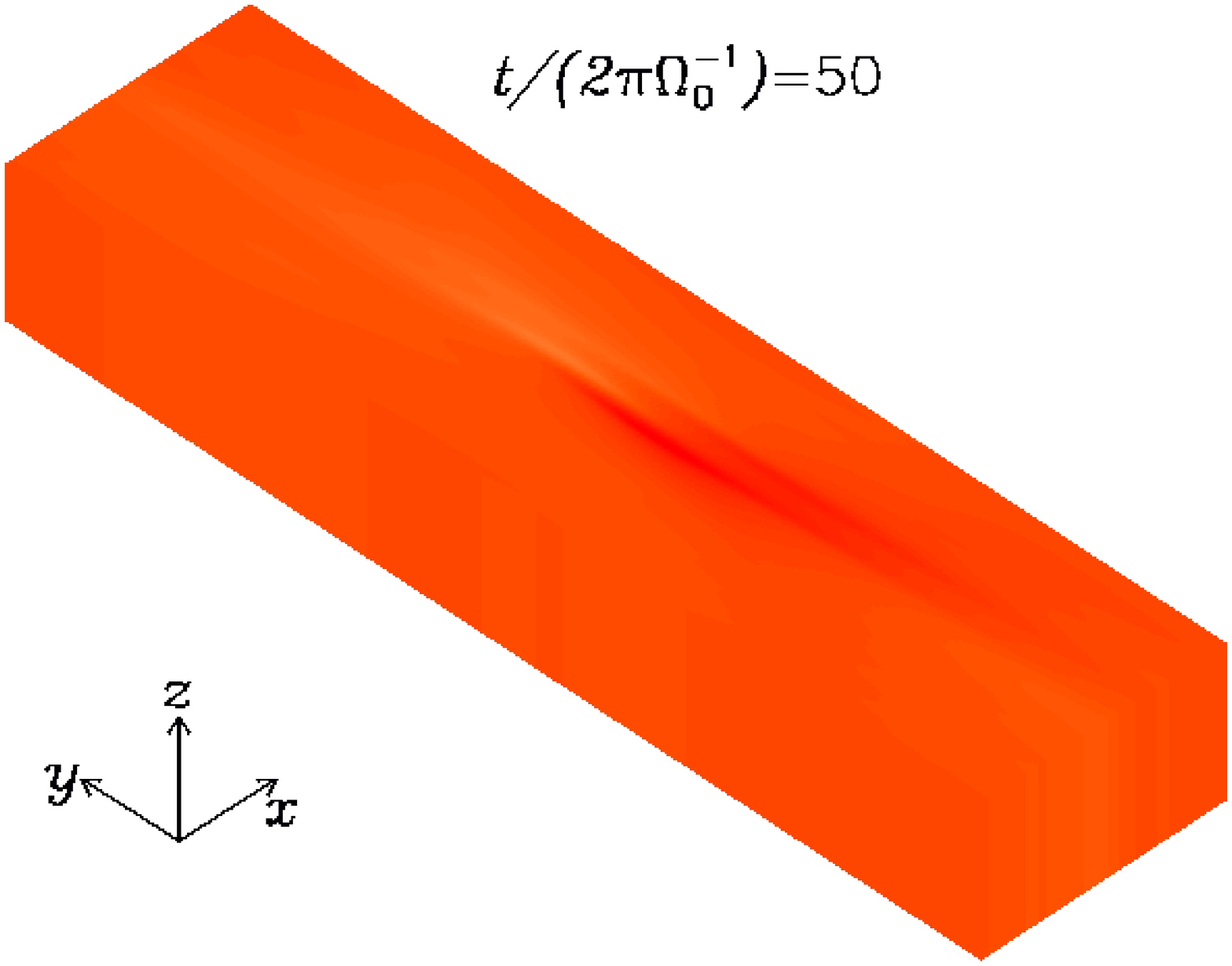}}
    \resizebox{.285\textwidth}{!}{\includegraphics{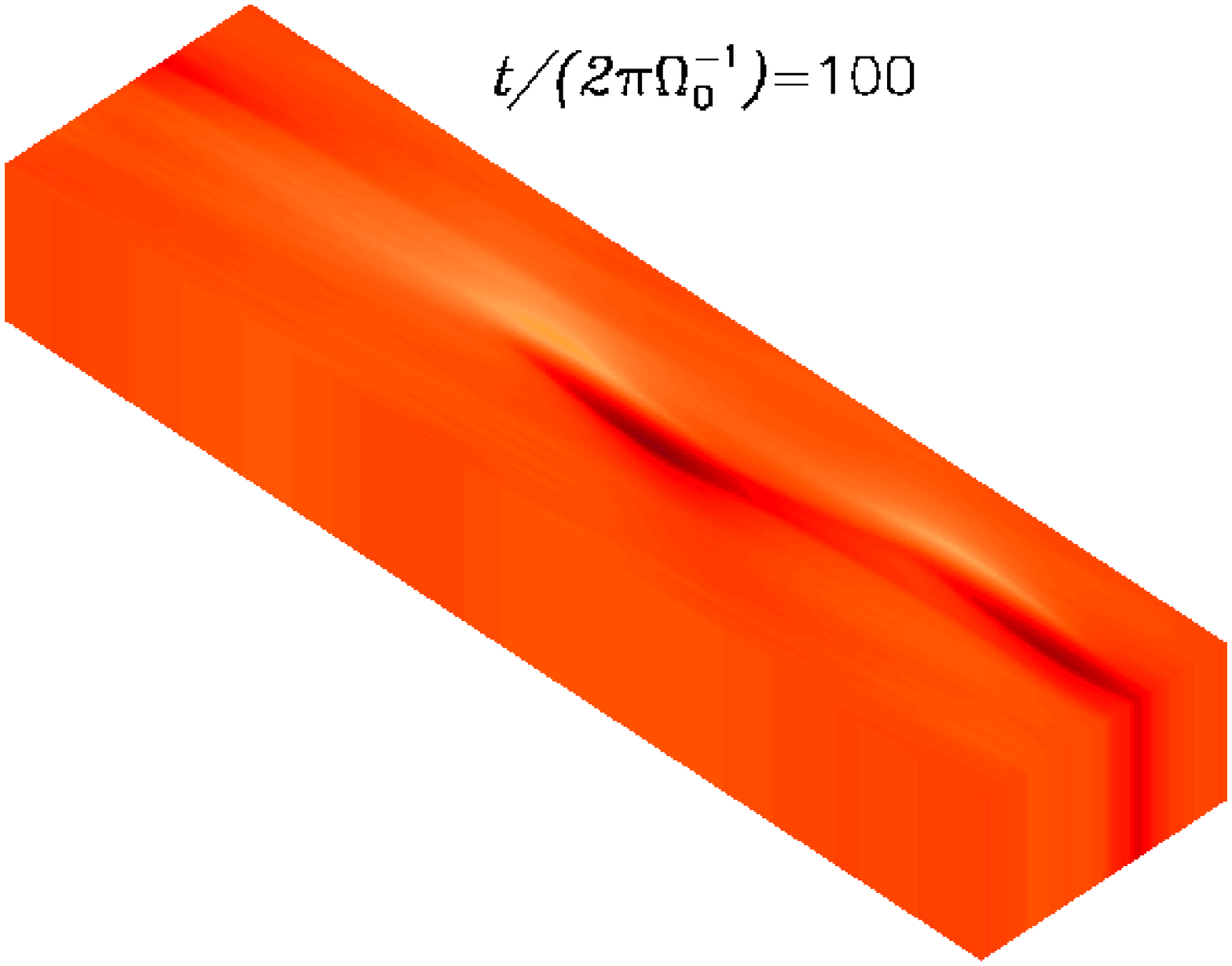}}
    \resizebox{.285\textwidth}{!}{\includegraphics{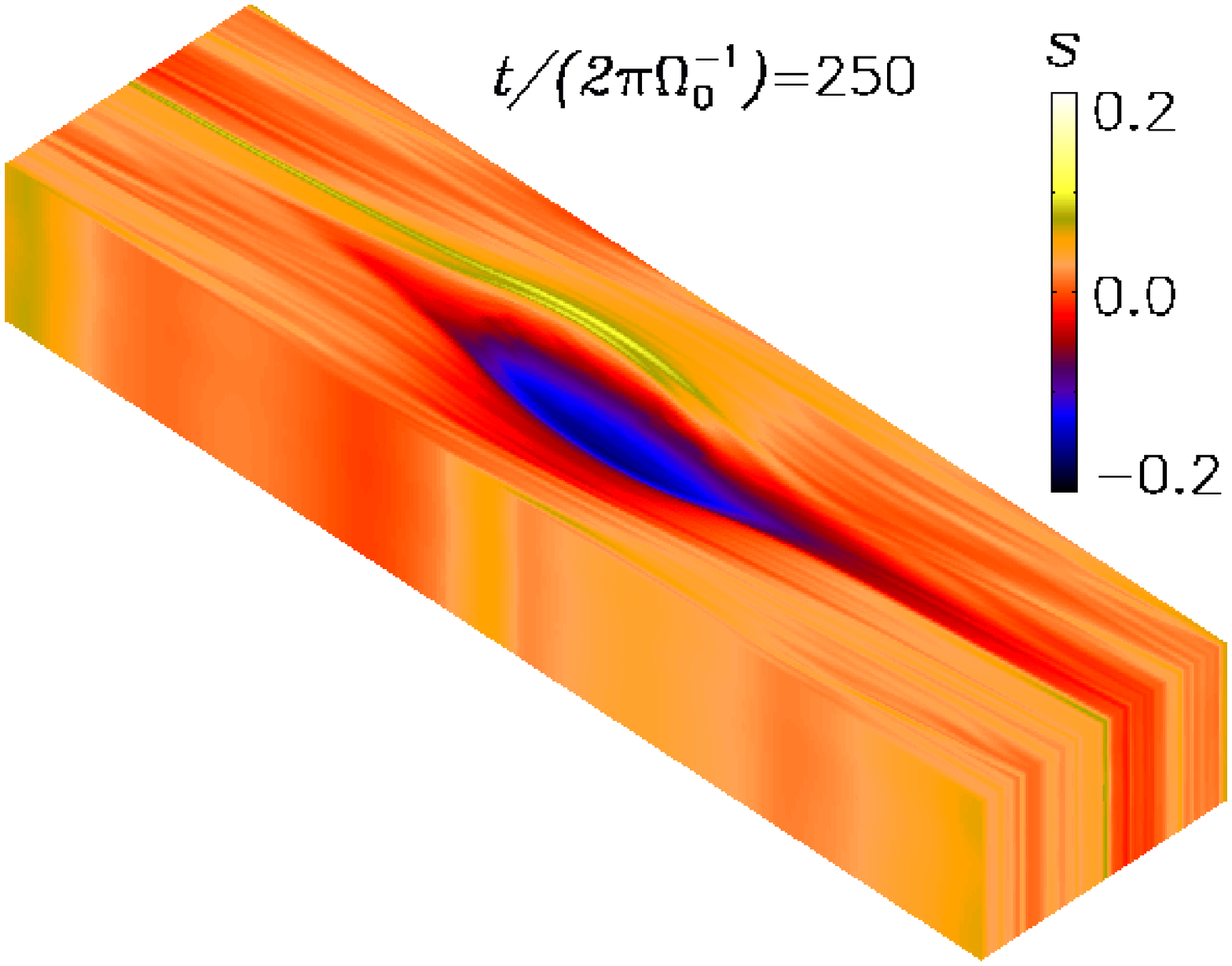}}
\end{center}
\caption[]{Evolution of vorticity (upper panels) and entropy (lower panels) due 
to the baroclinic instability in 3D.}
\label{fig:evolution-oo-ss}
\end{figure*}

\subsection{Diffusion}
\label{sect:Diffusion}

Thermal relaxation is one of the ways of changing the internal energy of a 
gas parcel. Another way is of course diffusion. Petersen et al. (2007a) and 
Lesur \& Papaloizou (2010) report sustained baroclinic growth using thermal diffusion, 
and this is also why the 3D simulations of Klahr \& Bodenheimer (2003) with 
flux-limited diffusion also experienced baroclinic growth. 
In the 2D simulations by Klahr \& Bodenheimer (2003), the thermal relaxation was 
numerical and a result of low resolution and a dispersive numerical scheme. 

We assess the effect of diffusion by setting $\tau_c$=$\infty$ (i.e., shutting down 
the thermal relaxation) and adding a non-zero radiative conductivity $K$ to the 
entropy equation (\eq{eq:entropy}). The thermal diffusivity $k_{\rm th}$, is related 
to the radiative conductivity by $k_{\rm th} \equiv K/(c_p\rho)$. As we choose dimensionless 
units such that $c_p$=$\rho_0$=1, then $k_{\rm th}\approx K$ in our simulations. The 
thermal diffusivity, 
like any diffusion coefficient, has a dimension of L$^2$T$^{-1}$, where L is length and T 
is time. We use L=$H$ where $H$=$c_s/\varOmega$ is the scale height, and write 
$k_{\rm th}$=$H^2/\tau_{\rm diff}$ so that the heat diffuses over one scale height within 
a time $\tau_{\rm diff}$. We assess $\tau_{\rm diff}$= 1, 10, and 100 orbits. These are 
runs F, G, and H in \Tab{table:hydro-disk-models}.
We see that now too fast a diffusion time (1 orbit) does not lead to growth, and 
vigorous growth occurs for 100 orbits. 
The rationale is the same as for the radiative case. Too fast diffusion disperses 
temperature gradients and weakens the baroclinic feedback. Slow diffusion works 
towards keeping the gradients tight and leads to vigorous growth. 

It is curious that the optimal diffusion time for growth is longer than in the 
thermal relaxation case. The difference between them is that relaxation 
is proportional to the temperature, whereas diffusion is proportional to the 
Laplacian of the temperature; that is, relaxation operates equally in all spectrum, 
while diffusion mostly affects higher frequencies. As such, stronger diffusion 
(when compared to relaxation) should be needed for longer wavelengths. At present, 
we can offer no explanation as to why this is not the case, though we notice that 
Lesur \& Papaloizou (2010) also see that the optimal diffusion time for the baroclinic 
feedback is substantially longer than the vortex turnover time. 

%
%
%
%

\subsection{Resolution}
\label{sect:Resolution}

To investigate the effect of resolution, we compare runs using 
128, 256, and 512 grid zones in the x-axis, and cells of unity aspect 
ratio (meaning four times more resolution in the y-axis than in the 
fiducial run). These are runs J, K, and L in \Tab{table:hydro-disk-models}.
They use the default values of $\tau_c$=1 orbit for the 
thermal relaxation time, and $\xi$=2 for the entropy gradient. 

As seen in \fig{fig:resolution}, the run with resolution 
128x512 (low resolution) fails to sustain vortex growth, in contrast 
to the runs with resolution 256x1024 (middle resolution) and 512x2048 
(high resolution). That the low-resolution run does not lead to enstrophy 
growth is a salutary reassurance that aliasing is not spuriously injecting 
vorticity in the box. The high-resolution run shows a slightly higher 
initial enstrophy production (from 0-30 orbits), yet it saturates to 
the same level as the middle resolution run, which suggests convergence. 

\subsection{Grid Reynolds number}
\label{sect:Grid-Reynolds-number}

As for the grid Reynolds number, we check three runs, with hyper-viscosities 
$\nu^{(3)}$=$\ttimes{-16}$ (run M), $\ttimes{-17}$ (run A), 
and $\ttimes{-18}$ (run N), 
meaning grid (hyper-)Reynolds numbers of $\xtimes{2}{-3}$, $\xtimes{2}{-2}$, 
and $\xtimes{2}{-1}$ with respect to the velocity shear introduced 
by the Keplerian flow, $\Rey = (3/2)\varOmega\,\Delta{x}^2/\nu$, where 
$\nu$=$\nu^{(3)}(\pi/\Delta{x})^4$.

By increasing the Reynolds number, the initial enstrophy 
amplification is stronger. Upon saturation, the mean enstrophy in 
all simulations converge to the same value. It should be noticed that 
although the grid Reynolds number upon saturation is greater than 1, 
the initial phase of growth occurs below this number - so growth 
cannot be due by aliasing. A heavily aliased solution 
is only attained when the hyperviscosity is decreased to $\nu^{(3)}$=$\ttimes{-21}$, 
so that the initial phase of growth occurs at very high Reynolds numbers 
(1000). At this Reynolds number, vortex growth occurs when the simulation 
is seeded with Gaussian noise, which is a sign that the growth was numerical, 
given the nonlinear nature of the baroclinic instability. In contrast, 
none of the simulations shown in the figure develop vortices when only seeded 
with noise. We usually use $\nu^{(3)}$=$\ttimes{-17}$, which yields a 
good compromise between not leading to aliasing and not affecting the 
timestep too much. 

\section{3D Results}
\label{sect:3D-Results}

Having examined the behavior of the baroclinic instability in 2D, we now 
turn to 3D simulations. We only study the unstratified case, because 
the stratified case needs a modification of the evolution equations, replacing $p_0$ in 
\Eqs{eq:navier-stokes}{eq:entropyeq} by $p_0f(z)$, where $f(z)$ is the stratification 
function. We use a box of length (4$\times$16$\times$2)\,${H}$, 
with resolution 256$\times$256$\times$128. Unlike in Lesur \& Papaloizou (2010), our 
simulations are compressible, which limits the timestep and makes it impractical 
to follow a 3D computation for many hundreds of orbits. For this reason, we follow it 
for 250 orbits, which was seen to be the beginning of saturation in 2D runs. The 
parameters of the simulation are shown in \Tab{table:hydro-disk-models} as run O.

\begin{figure}
  \begin{center}
    \resizebox{.9\hfwidthsingle}{!}{\includegraphics{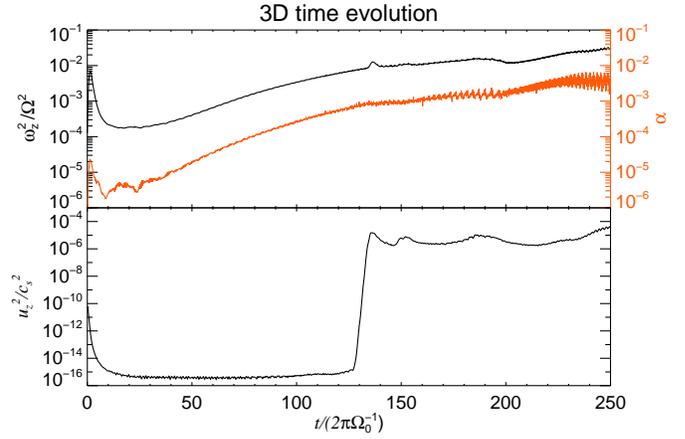}}
  \end{center}
\caption[]{Evolution of enstrophy, kinetic stresses, and vertical velocities in 
a 3D baroclinic simulation. The evolution is very similar to the 2D case 
up to 120 orbits. At that time the vortex goes elliptically unstable, and the kinetic 
energy of vertical motions increases by 10 orders of magnitude in less than 10 orbits, 
but remains three orders of magnitude lower than the radial rms velocity.
This 3D elliptical turbulence is very subsonic, and the vortex is not destroyed. The 
level of enstrophy and angular momentum transport remain similar to that of a 2D 
simulation.}
\label{fig:3d-timeseries}
\end{figure}

In \fig{fig:evolution-oo-ss} we show snapshots of enstrophy and entropy, and in 
\fig{fig:3d-timeseries} we plot the time series of box-averaged enstrophy, alpha value, 
and rms vertical velocity. As seen from these figures, the 3D baroclinic instability 
evolves very similarly to its 2D counterpart. After 200 orbits the instability 
begins to saturate as vortices merge and the remaining giant vortex grows to the 
sonic scale. The sandwich pattern of entropy perturbations sustaining the vortex is 
also very similar. The saturated state also displays similar values of enstrophy 
($\omega_z^2/\varOmega^2$ of the order of $\ttimes{-2}$) and angular momentum 
transport ($\alpha \approx \xtimes{5}{-3}$).

The difference is in the excitation of the elliptical instability 
(Kerswell 2002, Lesur \& Papaloizou 2009). As seen in the lower panel 
of \fig{fig:3d-timeseries}, 
the growth of this instability is very rapid, with the rms of the vertical 
velocity rising by ten orders of magnitude in less than 10 orbits. As in Lesur 
\& Papaloizou (2010), the instability leads to turbulence in the core of the 
vortex, but it is not powerful enough to break its coherence. This is because 
the elliptic destruction caused by the vortex stretching term is compensated 
with vorticity production by the baroclinic term. 

\begin{figure}
  \begin{center}
    \resizebox{.9\hfwidthsingle}{!}{\includegraphics{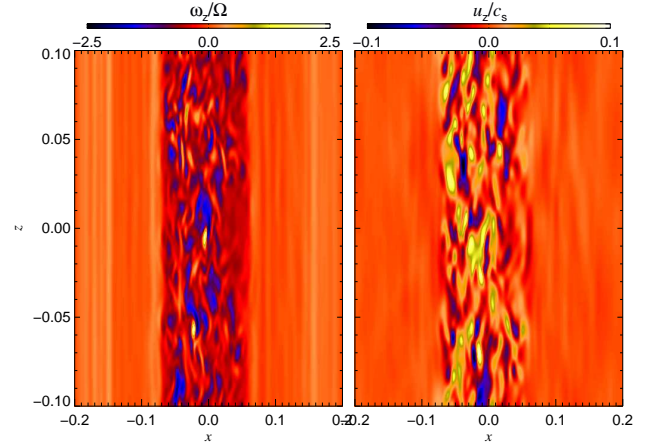}}
  \end{center}
\caption[]{Vertical slice of the elliptically unstable vortex core showing 
vertical vorticity (left panel) and vertical velocity (right panel). The motion 
in the core constitutes a subsonic turbulence at maximum speeds reaching 10\% 
that of sound.}
\label{fig:3d-vertical-slice}
\end{figure}

We follow the evolution of the vortex for 130 more orbits, without seeing 
any decay in the rms vertical velocity. In \fig{fig:3d-vertical-slice} 
we plot vertical slices of the $z$-vorticity and $z$-velocity, taken at the 
$y$-position of the vorticity minimum, at $t$=250 orbits. The snapshots reveal 
the vertical motions at the vortex core. The motion seems turbulent, only 
weakly compressible, with maximum velocities reaching 10\% of the sound speed.

We stress again that the alpha value is around $\xtimes{5}{-3}$ at saturation 
and positive. Lesur \& Papaloizou (2010) report a much lower (of the 
order of $\ttimes{-5}$) and negative angular momentum transport. This is 
because of the anelastic approximation, as the authors themselves point out. In 
that case, the angular momentum ``transport'' is solely due to the 3D 
instability that taps energy from the vortical motion. Compressibility allows 
for the excitation of spiral density waves, which enable positive angular 
momentum transport. 

\begin{figure}
  \begin{center}
    \resizebox{.9\hfwidthsingle}{!}{\includegraphics{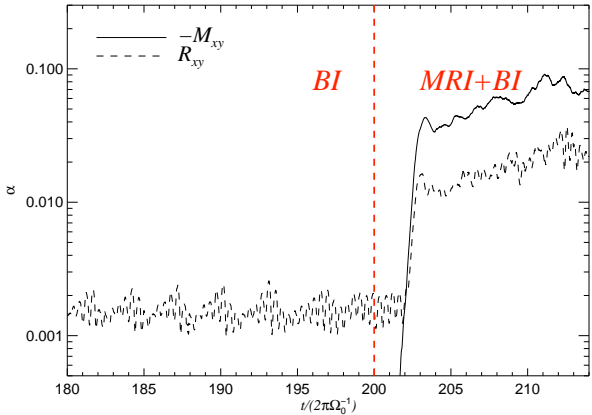}}
  \end{center}
  \caption[]{Angular momentum transport with only the baroclinic instability in a 
    3D run and the baroclinic and magneto-rotational instabilities, after 200 orbits. 
    The pattern after including the magnetic field is equal to that generated 
    by MRI-only, from which we conclude that the BI is irrelevant if magnetic fields 
    are well-coupled to the gas.}
  \label{fig:evolution-mribi-3b}
\end{figure}

\subsection{Magnetic fields. Interaction of the MRI and the BI}
\label{sect:Magnetic-fields}

The baroclinic instability demonstrated in the past sections seems to be able 
to drive angular momentum transport in accretion disks. As such, it could be 
thought of as an alternative to the MRI. Nevertheless, an important question to ask is 
how the two instabilities interplay. What happens if a magnetic field is 
introduced into the simulation? 

To answer this question, we take a snapshot of the quantities at 200 orbits, 
and add a constant vertical magnetic field to it, of strength 
$B$=$\xtimes{5}{-3}$ ($\beta$=$2\gamma^{-1} c_s^2/\va^2$$\approx$570). 
We assume ideal MHD, i.e.\ perfectly coupling of the field to the gas 
(run P in \Tab{table:magnetic-disk-models}). The same setup in a barotropic 
box leads to MRI turbulence with alpha values of the order of 
$\alpha\approx\xtimes{5}{-2}$. When the field is introduced into 
the MRI unstable box, the Maxwell stress immediately starts to grow, saturating after 
$\approx$3 orbits, as expected from the MRI. The Reynolds stress due to 
the MRI supersedes the stresses due to the BI by one order of magnitude 
(see \fig{fig:evolution-mribi-3b}). The 
pattern is the same as with an MRI-only box. The conclusion is immediate: the 
BI plays little or no role in the angular momentum transport when magnetic 
fields are well-coupled to the gas. This was intuitively expected, since the BI has weak 
angular momentum transport, as well as slow growth rates. The MRI is faster 
by 1 order of magnitude and much stronger. 

\begin{figure*}
  \begin{center}
    \resizebox{.9\textwidth}{!}{\includegraphics{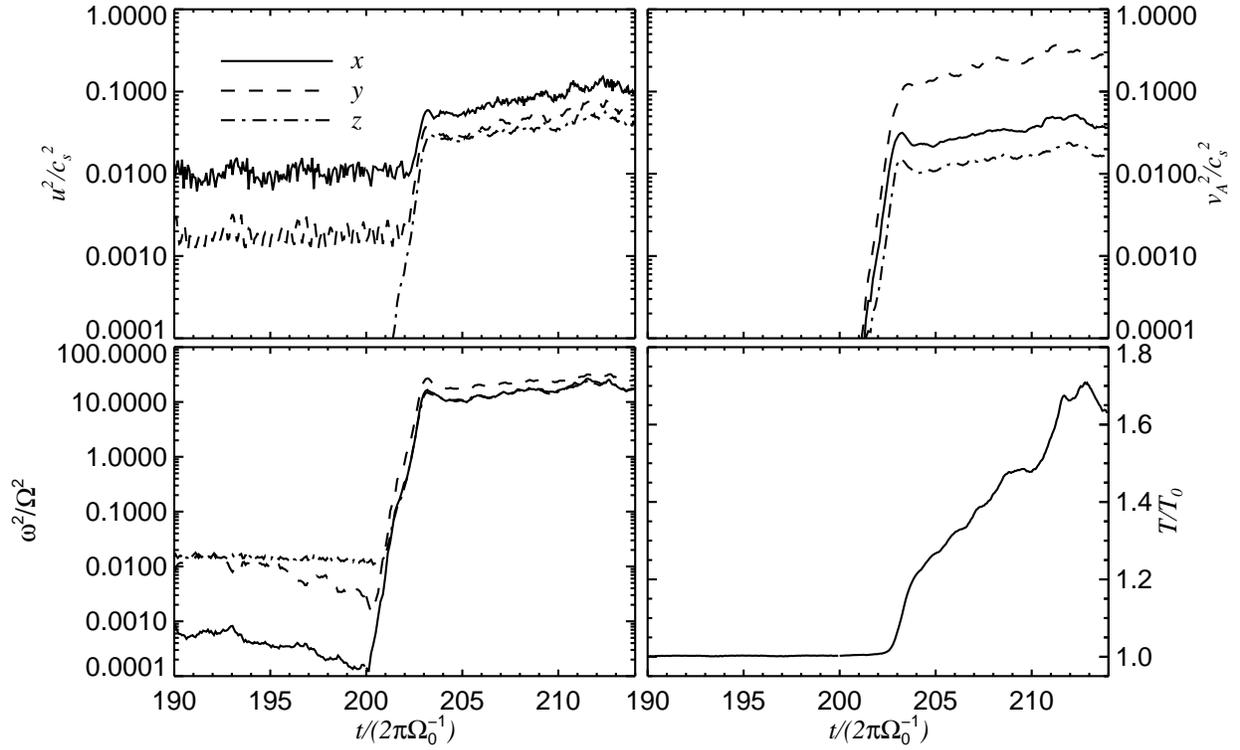}}
  \end{center}
  \caption[]{Evolution of box-average quantities (clockwise: kinetic 
energy, magnetic energy, enstrophy and temperature) before and after inserting 
the magnetic field. The MRI quickly takes over, on its characteristic 
short timescale. No evidence of synergy between the two instabilities is 
observed. The saturated state of the combined baroclinic+MRI resembles an 
MRI-only scenario.}
  \label{fig:evolution-mribi-boxaverage}
\end{figure*}

In \fig{fig:evolution-mribi-boxaverage} we
show the evolution of energies, enstrophy, and temperature, before and after 
including the magnetic field (at 200 orbits). The magnetic energy behaves as 
expected from the MRI, a fast growth and saturation after $\approx$3 orbits, 
with most of the energy stored in the azimuthal field. The kinetic energy 
of the turbulence increases by one order of magnitude and is more isotropic, also as 
expected from the MRI. The temperature increases by a factor of $\approx$2 in 
15 orbits. This is because the MRI turbulence heated the box faster than the thermal 
relaxation time could bring the temperature back to $T_0$.

With this experiment we expected to assess the possibility of synergy between 
the instabilities, but as far as we can tell, none is observed because the MRI alone 
dictates the evolution. 

\begin{figure*}
  \begin{center}
    \resizebox{.285\textwidth}{!}{\includegraphics{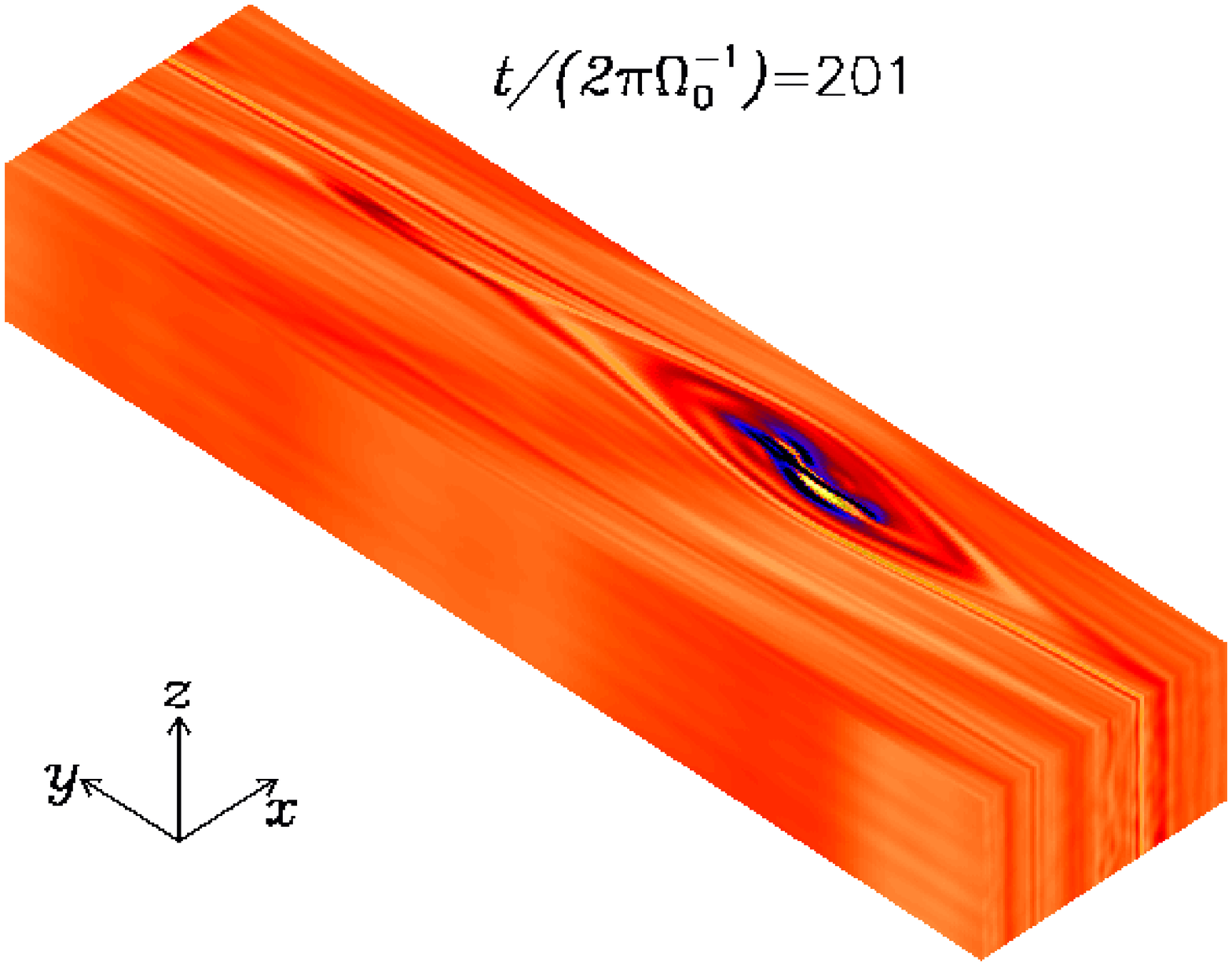}}
    \resizebox{.285\textwidth}{!}{\includegraphics{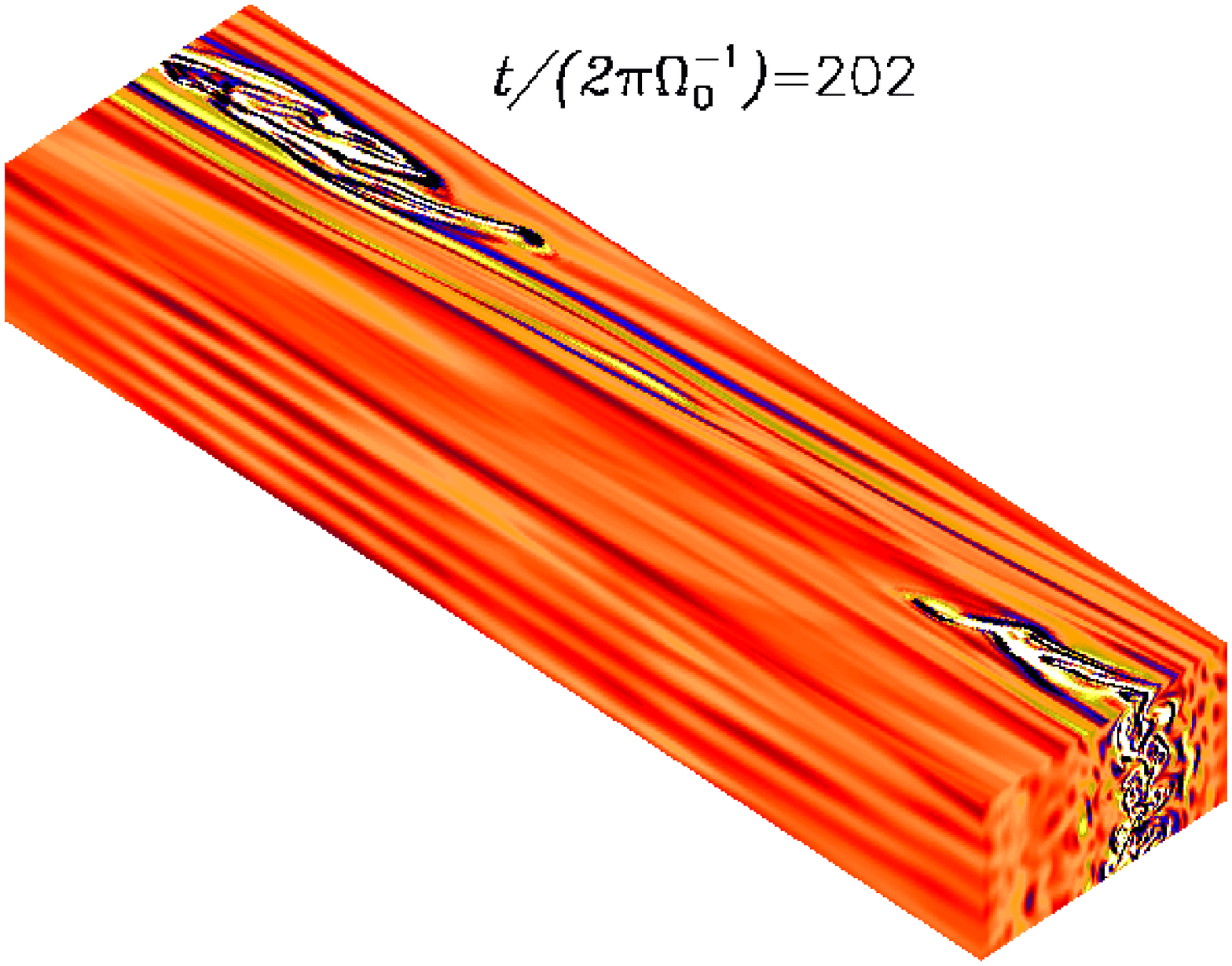}}
    \resizebox{.285\textwidth}{!}{\includegraphics{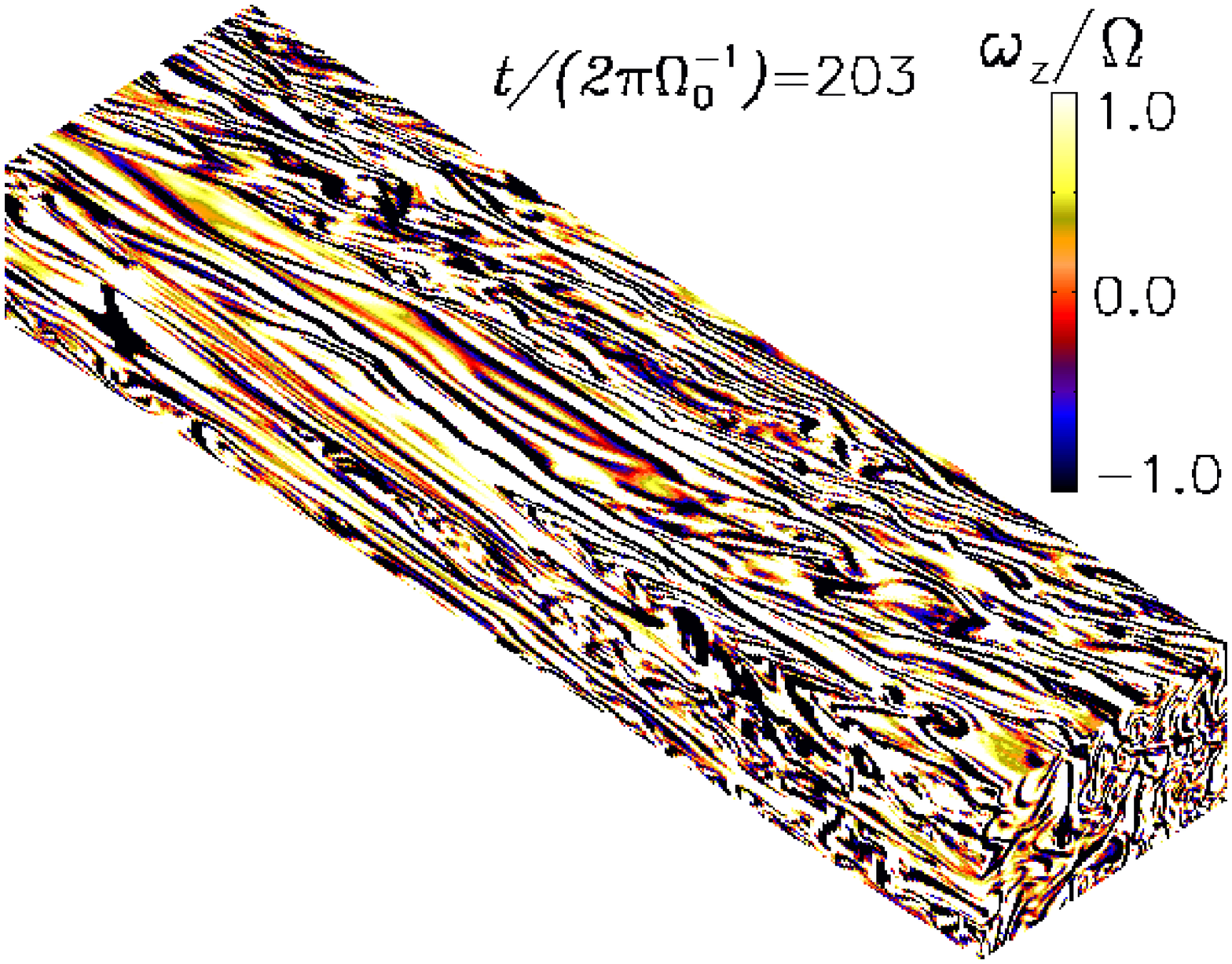}}
    \resizebox{.285\textwidth}{!}{\includegraphics{figs/15568fg11d.ps}}
    \resizebox{.285\textwidth}{!}{\includegraphics{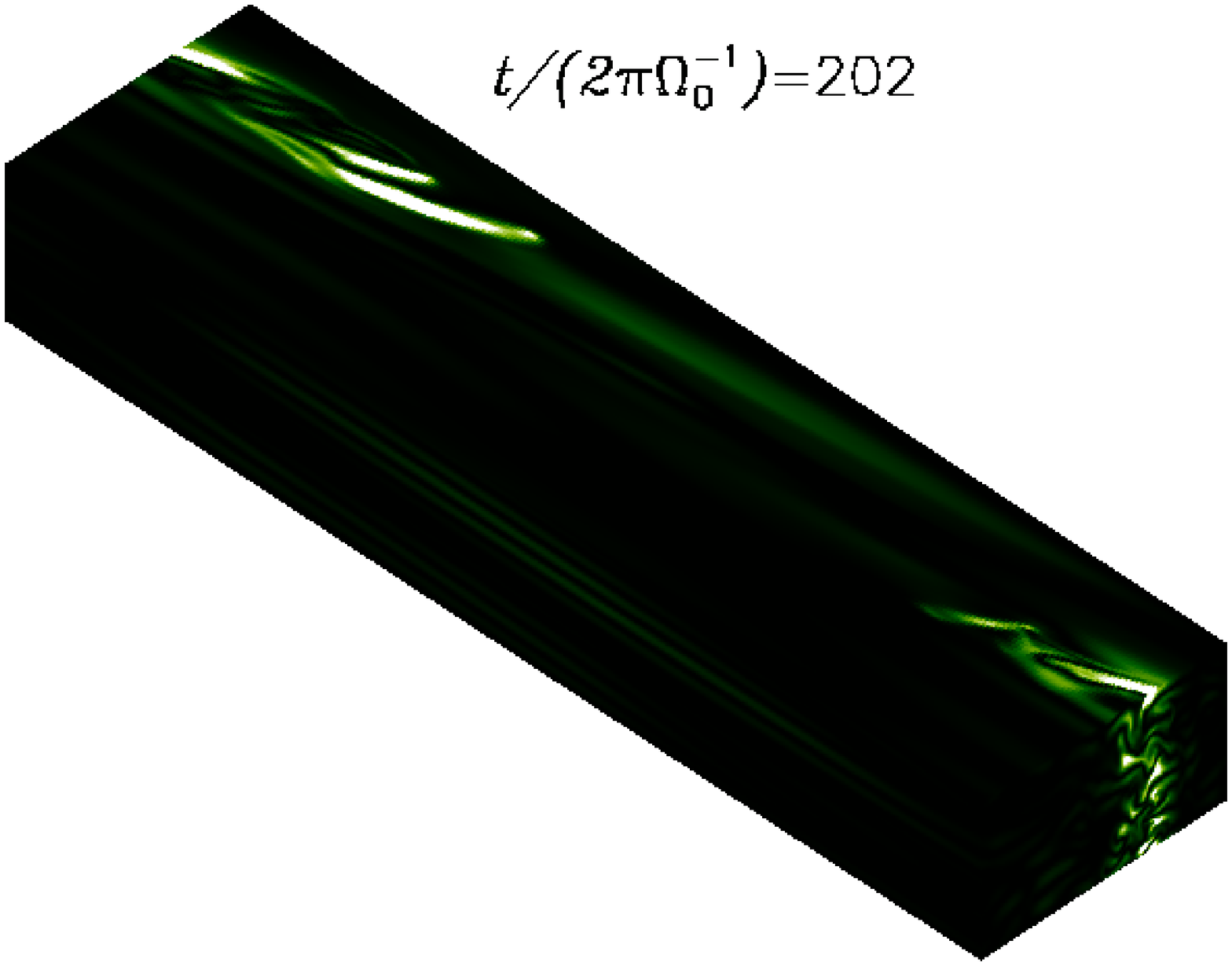}}
    \resizebox{.285\textwidth}{!}{\includegraphics{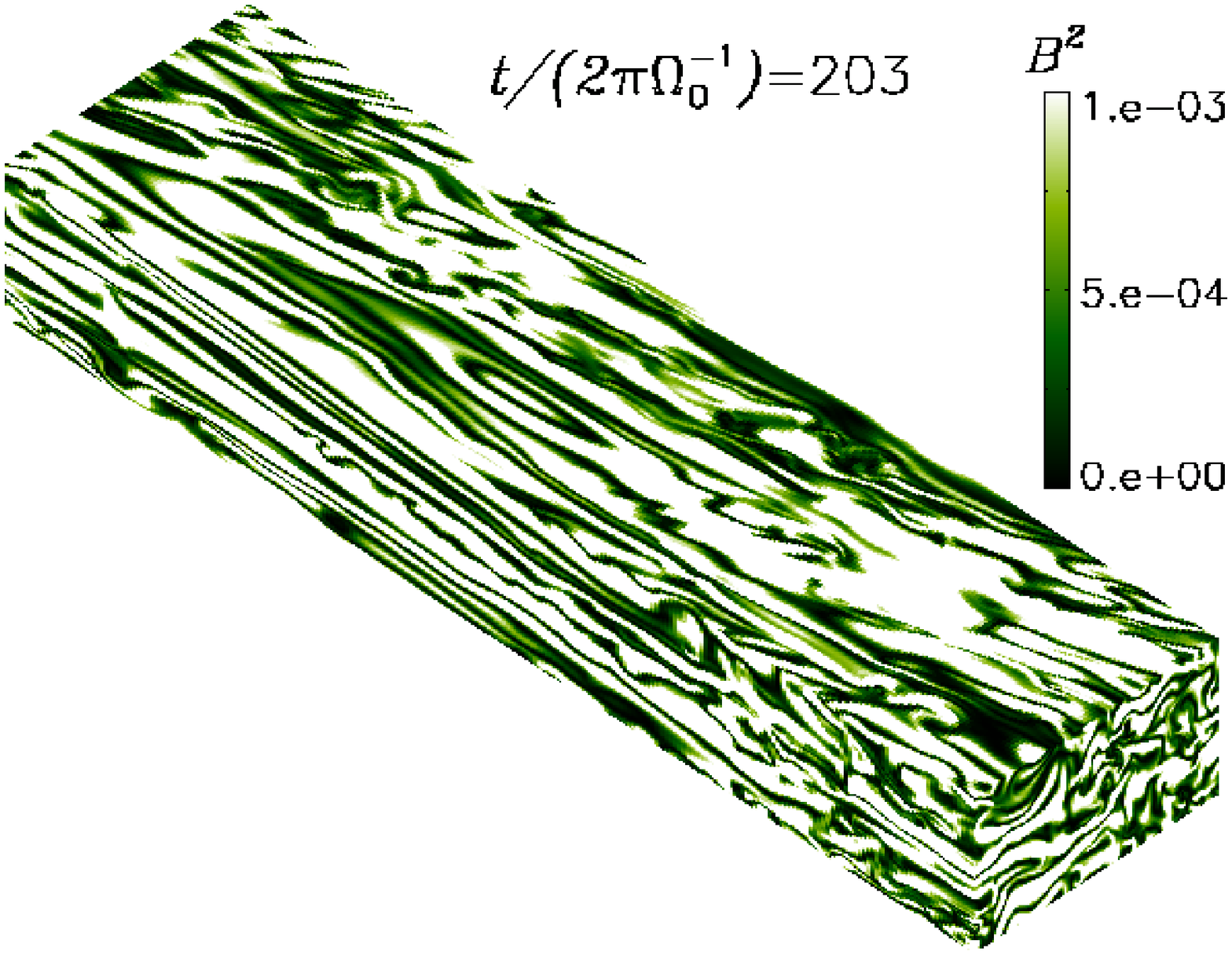}}
\end{center}
\caption[]{Evolution of vorticity (upper panels) and magnetic energy 
(lower panels) in 3D. As the MRI develops, the vortex 
is destroyed by the magnetic field. In a nonmagnetic run, the vortex 
survives indefinitely.}
\label{fig:evolution-oo-bb}
\end{figure*}

\begin{figure*}
  \begin{center}
    \resizebox{.76\textwidth}{!}{\includegraphics{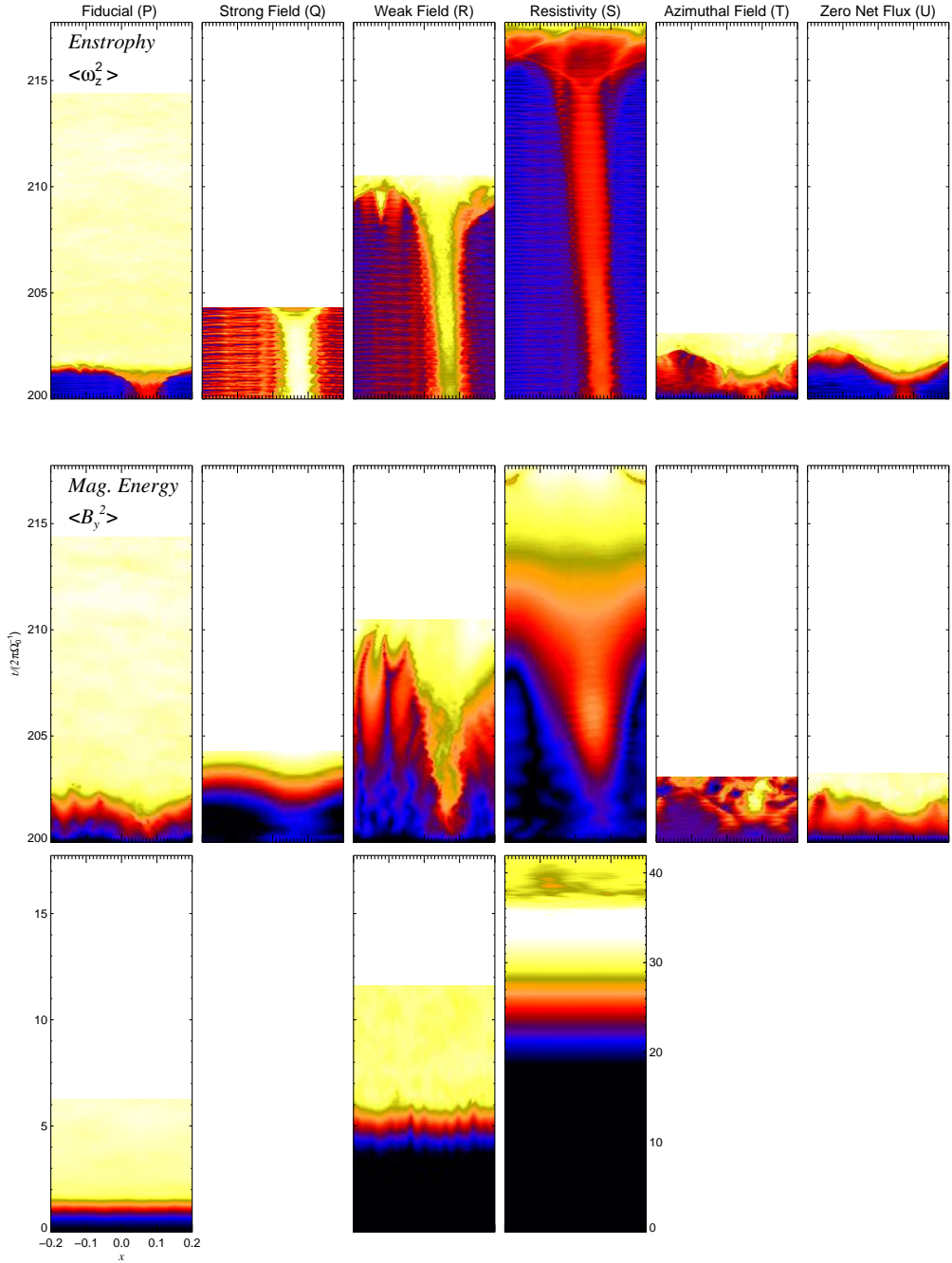}}
\end{center}
\caption[]{Time evolution of the 1D spatial average of enstrophy (upper 
panels) and azimuthal magnetic field (middle panels) for the runs in 
\Tab{table:magnetic-disk-models}. The lower panels refer to the magnetic field 
attained in control runs, where $\ksi$=0. In all runs, the field is seen to 
grow first in the vortex, then in the surrounding flow. This shows that the 
growth rates of the magneto-elliptic instability are faster than those of the MRI. 
Vortex destruction is apparent in these plots as loss of spatial coherence in 
the enstrophy plots, and occurred in all simulations. The length of the 
time axis is the same for all simulations, except the control run for run S.}
\label{fig:spacetime}
\end{figure*}

\begin{figure*}
  \begin{center}
    \resizebox{.85\textwidth}{!}{\includegraphics{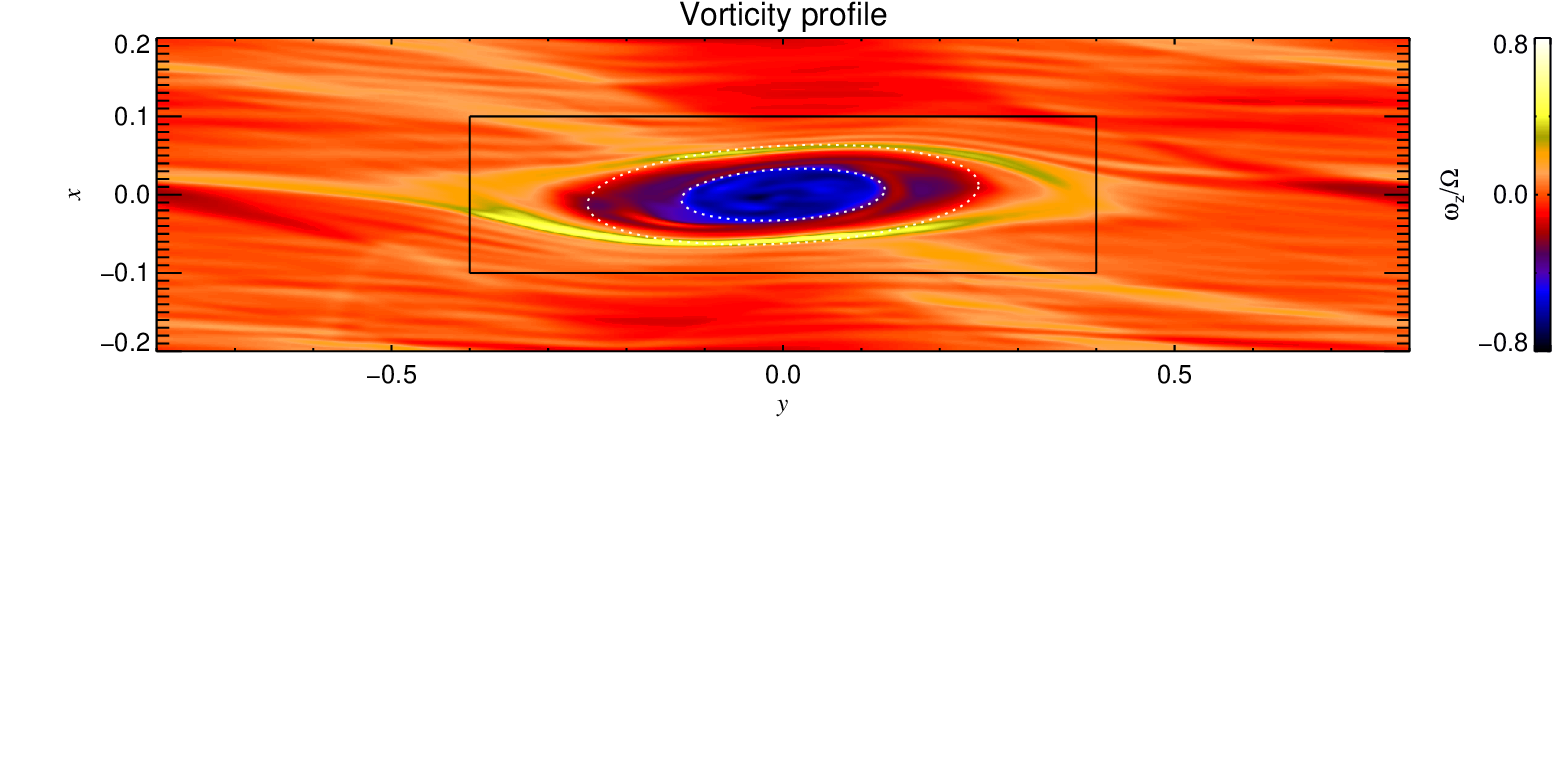}}
\end{center}
\caption[]{Vorticity profile of the vortex, prior to the insertion of 
the magnetic field. We measure the vertical vorticity in the midplane of the 
simulation against the elliptical radius, in the grid points boxed by 
the thin black line as shown in the upper panel. The modulus of the 
vorticity is plotted in the lower panel. The conclusion is that the vortex 
core has an angular velocity profile close to uniform, with shear 
where it couples to the Keplerian flow. The dashed lines in the lower 
panel mark the position of the dotted ellipses in the upper panel. They 
have an aspect ratio $\chi$=$4$, and mark elliptical radii of $r_V$=0.065 
and $r_V$=0.13. The inner one encloses the vortex core.}
\label{fig:vortex-shear}
\end{figure*}

\subsubsection{Vortex destruction by the magnetic field}
\label{sect:Vortex-destruction-by-the-magnetic-field}

If the evolution of the box-averaged quantities brings no surprises, the 
same cannot be said of their spatial distribution. In \fig{fig:evolution-oo-bb} we 
plot vorticity at three consecutive orbits after insertion of the magnetic field. 
Magnetic energy is also shown. The vortex, which in a nonmagnetic run retains its 
coherence indefinitely, is dilacerated when magnetic fields are included. In 
\fig{fig:spacetime} we plot 1D spatial averages against time of the vertical 
enstrophy (upper panel) and magnetic energy of the azimuthal field (middle panel). 
A control run where $\ksi=0$ is also shown (lower panel). The figure also shows other 
simulations (discussed later). The run in question is the leftmost one, 
labeled ``P''. It is apparent from the enstrophy plot that the vortex bulges, 
then gets destroyed as the magnetic energy grows.

To understand this behavior, we examined the state of the vortex prior to 
inserting the field. In \fig{fig:vortex-shear} we measure the vorticity profile 
of the vortex. The figure shows a slice at the midplane, where we define a box of 
size 8$H$ $\times$ 2$H$ centered 
on the vorticity minimum. We used elliptical coordinates such that the 
radius is $r_V$=$(\chi x_c^2 + y_c^2/\chi)^{1/2}$, where $\chi=a/b$ is the 
vortex aspect ratio ($a$ is the semimajor and $b$ the semiminor axis). The coordinates 
$x_c$ and $y_c$ are rotated by a small angle to account for the off-axis tilt of the vortex, 
$(x_c,y_c)$=$\vt{R}(x-x_0,y-y_0)$, where $\vt{R}$ is the rotation matrix, and $(x_0,y_0)$ 
are the coordinates of the vortex center, found by plotting ellipses and looking for a 
best fit. We find that $\chi$=4 and a rotation of 3$^\circ$ best fits the vortex core. 
Two such ellipses are plotted in the upper panel of \fig{fig:vortex-shear}.

We then measured the vertical vorticity within the box, averaged 
all vertical measurements for a given radius, and box-plotted the 
$z$-averaged measurements against $r_V$. The box plot 
uses a bin of $\Delta{r_V}=0.01$. The result is seen in the lower panel of 
\fig{fig:vortex-shear}, with the radius of the ellipses 
drawn in the upper panel. It is seen that the vortex core 
(inside the inner ellipse) has a vorticity profile that is well 
approximated by a Gaussian,  
$\omega_V = \omega_0 \, {\rm exp}(-r_v^2/2r_0^2)$, where $\omega_0=0.62$, $r_0=0.1$, 
and the radii are in elliptical coordinates. 

We conclude that the vorticity in the core is close to uniform 
(as a Gaussian is very flat near the peak amplitude). Because the vorticity 
is finite and close to uniform, so is the angular momentum, and thus 
little radial shear should be present in the core. As the MRI 
feeds on shear, one can expect that a patch 
of constant (or nearly constant) angular momentum should be stable. 
Nevertheless, examining the vorticity after 2 orbits of the insertion 
of the field, (upper middle panel of \fig{fig:evolution-oo-bb}), we 
notice that the core did become unstable. This seems to 
be a signature of the magneto-elliptic instability (Lebovitz \& Zweibel 2004, 
Mizerski \& Bajer 2009), which we consider in the next section.

\subsubsection{Magneto-elliptic instability}
\label{sect:Magneto-elliptic-instability}

The elliptic instability has been a topic of extensive study in 
fluid mechanics (see review by Kerswell, 2002). First studied in 
the context of absent background rotation (Bayly 1986, Pierrehumbert 1986), 
the effect of the Coriolis force was studied by Miyakazi (1993), followed 
by the effect of magnetic fields by Lebovitz \& Zweibel (2004). The general 
case, in which both background rotation and magnetic fields are present, has 
recently been studied by Mizerski \& Bajer (2009). 

These studies have unveiled two regimes of operation, which may as well 
be seen as two different instabilities. The first de-stabilization 
mechanism is through resonances between the frequency of inertial waves and 
harmonics of the vortex turn-over frequency. This instability is three-dimensional, 
existing for $\theta > 0$ (the angle $\theta$ being the angle between the 
wavevector of the pertubations and that of the vortex motion). 
Lebovitz \& Zweibel (2004) show that this instability persists in the 
presence of magnetic fields, and that its effect is twofold. While it lowers 
the growth rates of the elliptically unstable modes, the excitation of 
MHD waves allows for de-stabilization of whole new families of resonances. 

The second destabilizing mechanism occurs only when the Coriolis force is included 
(Miyakazi 1993). This instability is nonresonant in nature and exists only for 
$\theta$=0 modes, i.e., oscillations in the same plane of the motion of the 
vortex. Because this plane is associated with a ``horizontal'' (xy) 
plane (thus $k_z$ modes), this destabilizing mechanism has been 
called ``horizontal instability''. As shown by Lesur \& Papaloizou (2009), 
this nonresonant instability results in exponential drift of epicyclic 
disturbances. It can thus be regarded as an analog of the Rayleigh instability, 
but for elliptical streamlines. For a vortex embedded in a Keplerian disk, 
the modified epicylic frequency goes unstable for the range of aspect 
ratios 3/2$<$$\chi$$<$4. 

Mizerski \& Bajer (2009) present the analysis of the general case, 
including both the Coriolis and Lorentz forces. They confirm the 
previous effects of the Coriolis and Lorentz forces in isolation, and 
find that the horizontal instability, when present, dominates over all 
other modes. They also find that the magnetic field widens the range 
of existence of the horizontal instability to an unbounded interval of 
aspect ratios when

\begin{figure}
  \begin{center}
    \resizebox{.45\hfwidth}{!}{\includegraphics{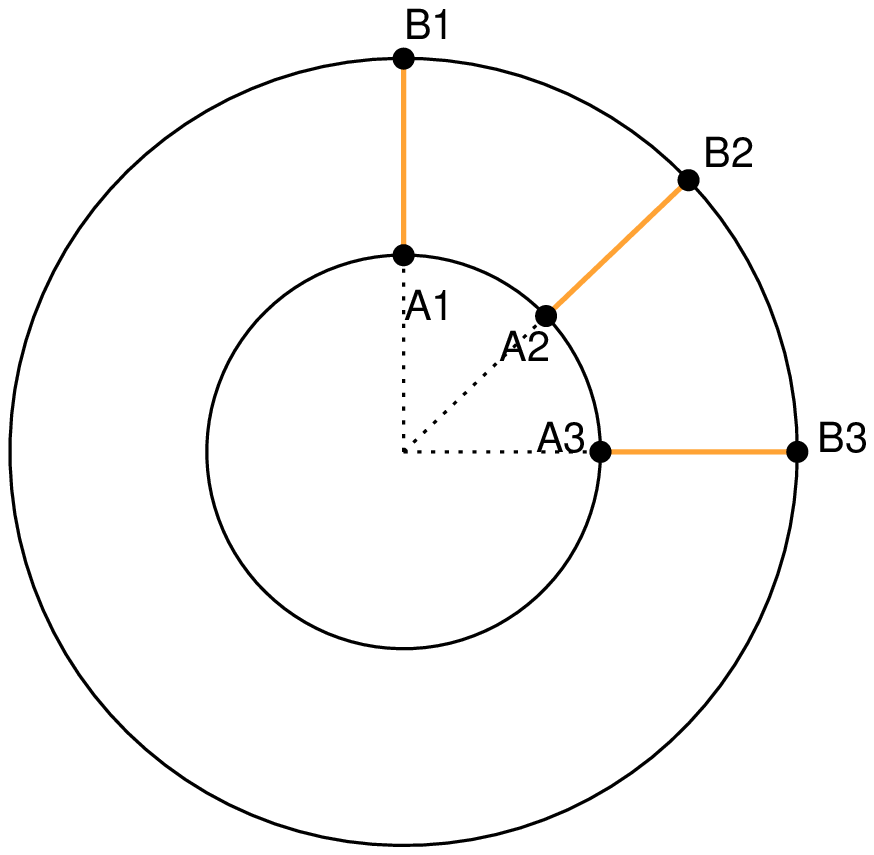}}
    \resizebox{.9\hfwidth}{!}{\includegraphics{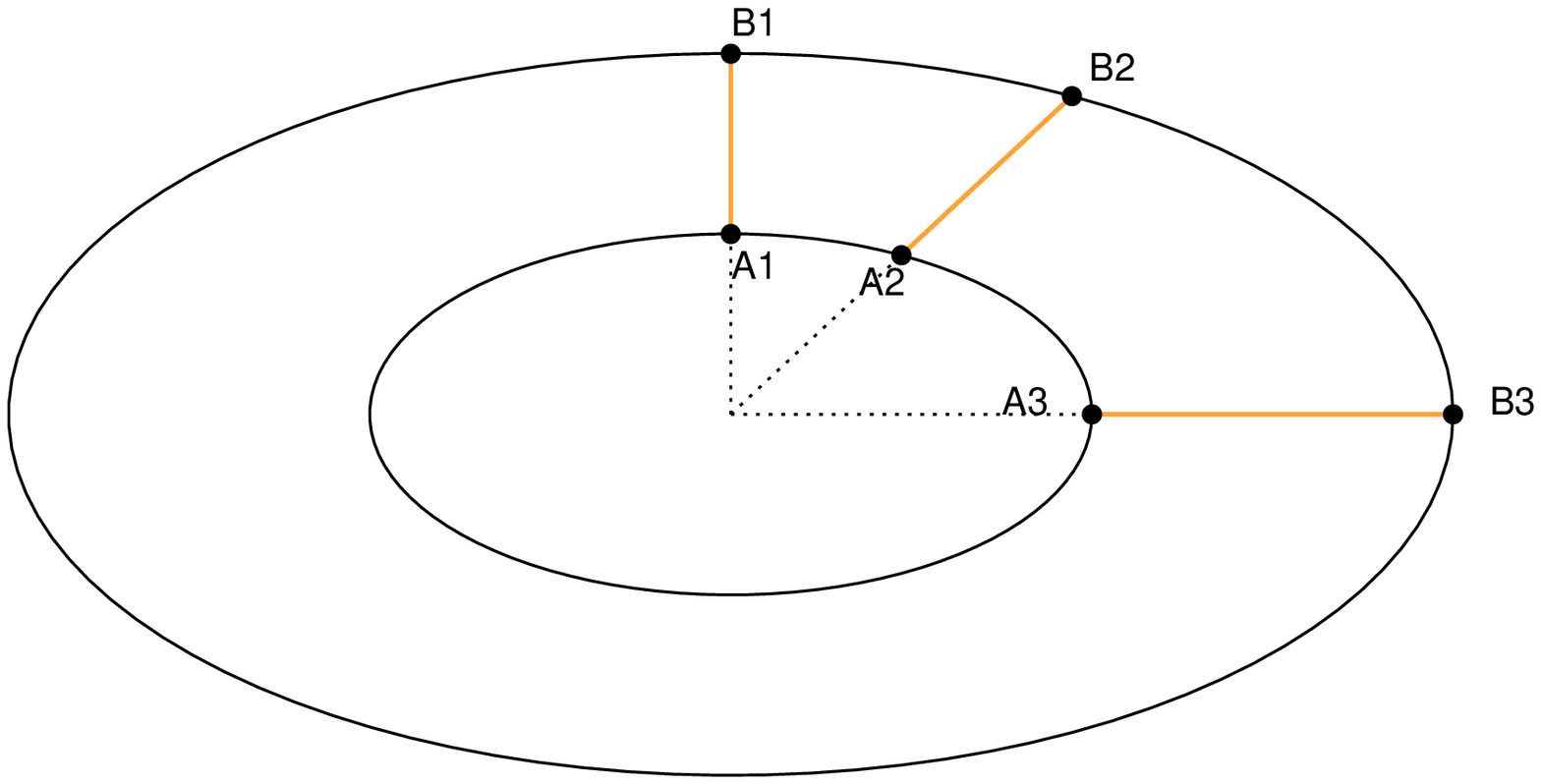}}
\end{center}
\caption[]{Shear as the common ground between 
the magneto-rotational and magneto-elliptical instabilities. The distance 
between two points in uniform rotation does not increase if the streamlines 
are circular, i.e., the rotation is rigid (upper figure). However, in elliptic 
streamlines the distance between the two points does increase even if the 
rotation is uniform (lower figure). A magnetic field connecting the two points 
will resist this shear, leading to instability depending on the field strength.}  
\label{fig:circle-ellipse}
\end{figure}

\begin{figure*}
  \begin{center}
    \resizebox{.9\hfwidth}{!}{\includegraphics{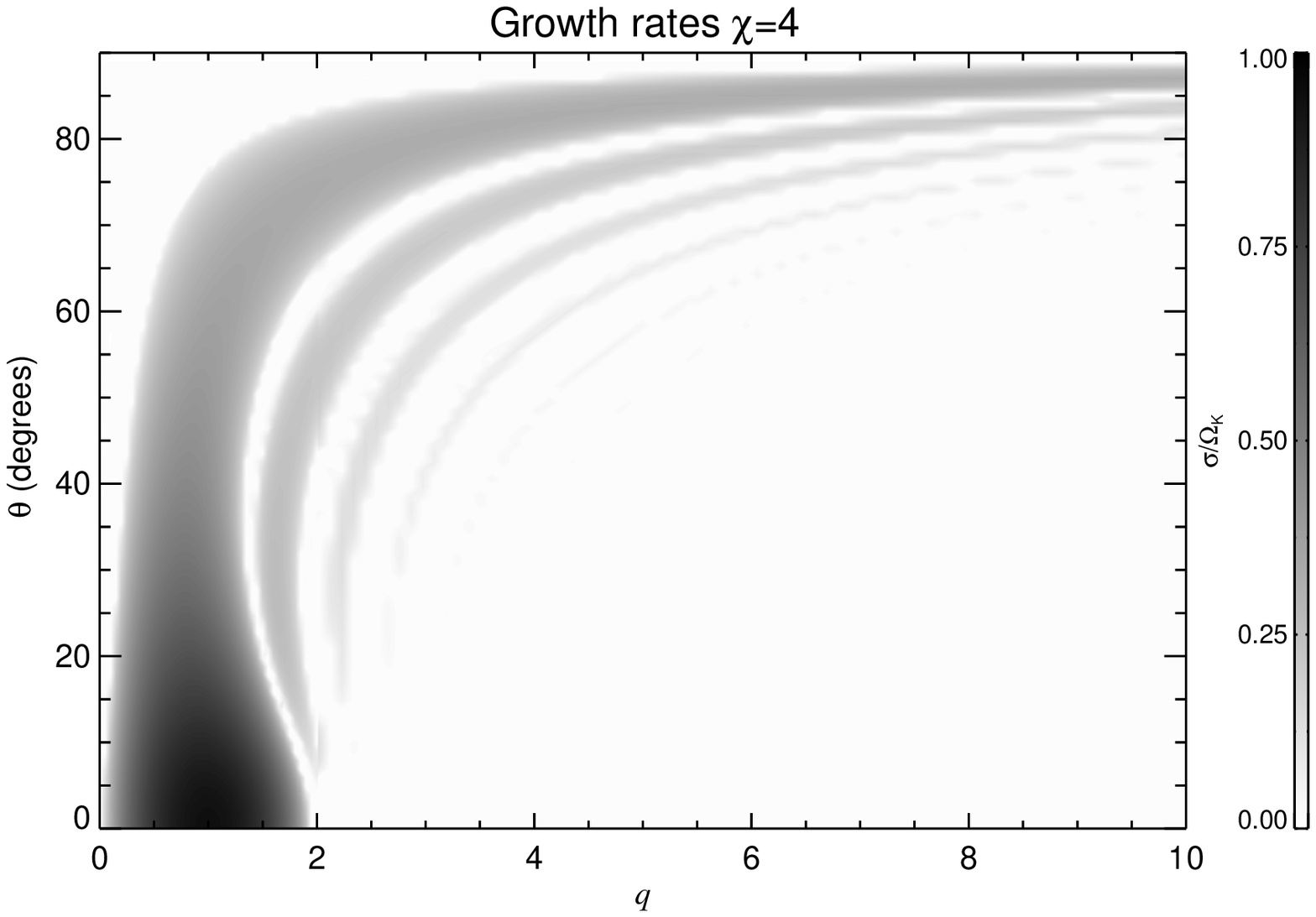}}
    \resizebox{.9\hfwidth}{!}{\includegraphics{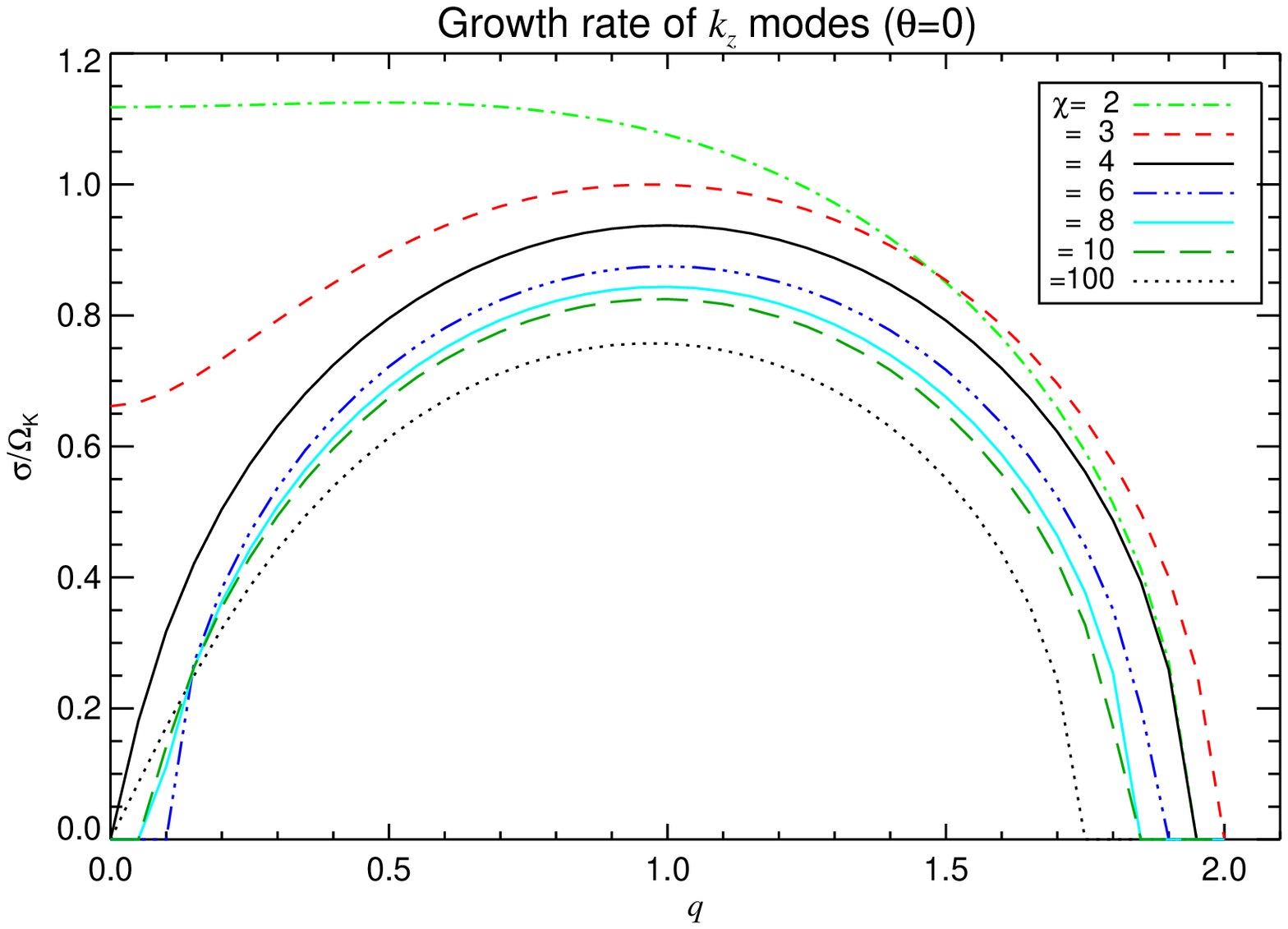}}
\end{center}
\caption[]{{\it Left.} Numerically calculated growth rates of the magneto-elliptic 
instability for elliptic streamlines of aspect ratio $\chi=4$ as a function of 
the dimensionless magnetic field strength $q$=$k\va/\varOmega_K$ and the angle 
$\theta$ between the wavevector of disturbances and the vertical axis. Pure 
$k_z$ modes are the most unstable ones, having a critical wavelength near the 
predicted $q_{\rm crit}$=$2\sqrt{|\Ro|}$. Weaker destabilization exists at 
intermediate $\theta$ (3D disturbances) for shorter wavelengths. 
Pure planar disturbances ($\theta$=$\pi/2$) are stable.\\
{\it Right.} Growth rates of the $k_z$ modes for different aspect ratios. For 
$\chi=2$ and $\chi=3$ the purely hydrodynamical elliptical instability is seen 
as finite (and high) growth rates as $q$ tends to zero. For $\chi = 4$ onwards, 
the instability is magnetic and has a most unstable wavelength near $q=1$. The 
$\chi=100$ curve stands for an approach to the limit of pure shear flow. The 
growth rate curve calculated matches that of the MRI.} 
\label{fig:growth-rates}
\end{figure*}

\begin{figure*}
  \begin{center}
    \resizebox{.85\textwidth}{!}{\includegraphics{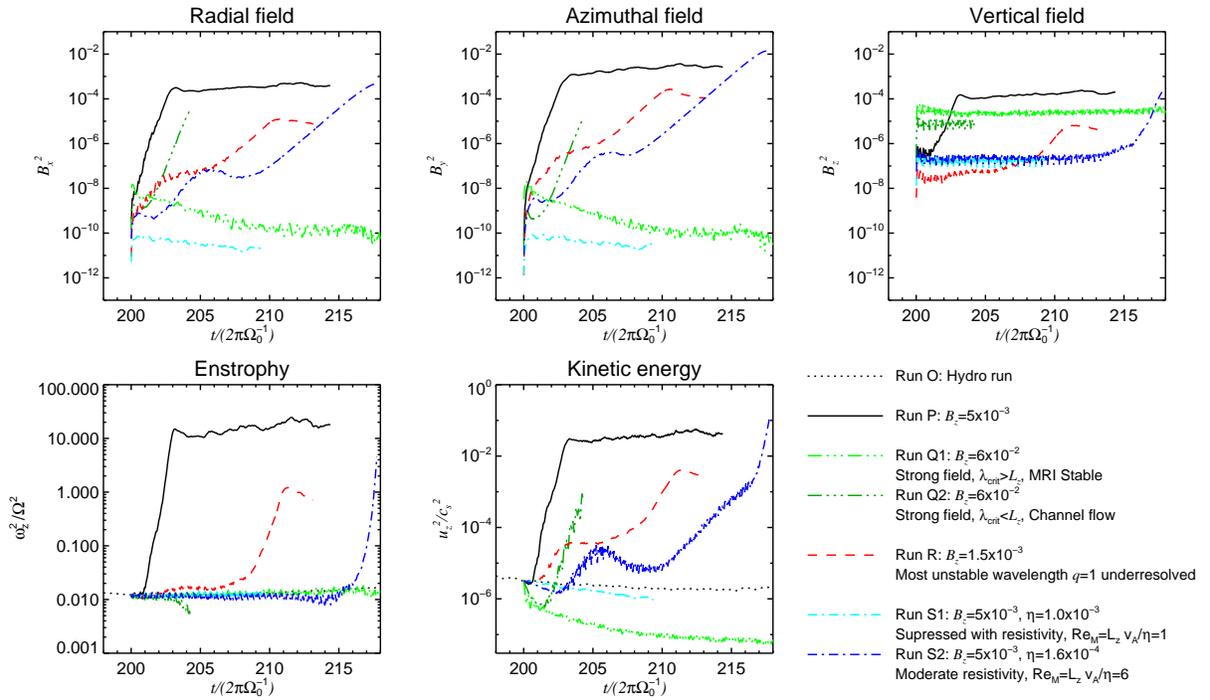}}
\end{center}
\caption[]{Isolating the vortex magnetic action. The lines show the magnetic runs with 
the MHD instabilities (magneto-elliptic and magneto-rotational) resolved in ideal MHD (P, black solid); 
unresolved with strong field (Q1, green dot-dot-dot-dashed); most unstable wavelength under-resolved 
with weak field (R, red dashed); and quenched with resistivity (S, blue dot-dashed). The nonmagnetic 3D 
hydro run is shown as dotted line in the lower panels. Run Q1 shows that strong magnetic fields have a 
stabilizing effect on the elliptical instability. Run R shows the magneto-elliptic instability seeming to 
saturate (at 203 orbits) before the MRI takes over (at 207 orbits). In runs Q2 and S2, the vortex survives until a 
channel flow develops in the box.}
\label{fig:magruns}
\end{figure*}

\begin{equation}
  \Ro^{-1} < -\frac{b^2}{4}, 
  \label{eq:magnetoelliptic}
\end{equation}where 

\begin{equation}
  \Ro=\frac{\varOmega_V\delta}{\varOmega_K} 
\end{equation}is the Rossby number,

\begin{equation}
  \delta=(\chi+\chi^{-1})/2
\end{equation}is another measure of the ellipticity, and 

\begin{equation}
  b=\frac{k\va}{\varOmega_V\delta} 
\end{equation}is a dimensionless parametrization of the magnetic field, 
with $k$ the wavenumber. We can also write $b$=$q/\Ro$, where 

\begin{equation}
  q=\frac{k\va}{\varOmega_K} 
\end{equation}is a more usual dimensionless parametrization of the 
magnetic field, based on the Balbus-Hawley wavelength 
$\lambda_{\rm BH}=2\pi\va/\varOmega_K$ (Balbus \& Hawley 1991, 
Hawley \& Balbus 1991). The analysis of Mizerski \& Bajer (2009) also 
assumes that the vortex is of the type 

\begin{eqnarray}
  u_x &=& -\varOmega_V (y-y_0)/\chi \nonumber\\
  u_y &=& \white{-}\varOmega_V (x-x_0)\chi. \label{eq:streamlines}
\end{eqnarray}

When there is a magnetic field but the criterion posed by \eq{eq:magnetoelliptic}
is not fulfilled, Mizerski \& Bajer (2009) find that 
the field has an overall stabilizing effect on the resonant modes of the classical 
(hydro) elliptic instability. 

We can rewrite \eq{eq:magnetoelliptic} in more familiar terms by isolating the 
wavenumber and expressing the criterion in terms of $\varOmega_K$ and $\Ro$

\begin{equation}
  0 < k < 2\,|\Ro|^{1/2}\,\frac{\varOmega_K}{\va}.
  \label{eq:criterion}
\end{equation}

We estimate the Rossby number of the vortex in \fig{fig:vortex-shear}. 
Assuming the elliptic flow of \eq{eq:streamlines}, 
the vorticity is $\omega_T = 2\delta\varOmega_V$. The subscript $T$ stands for ``total''. 
This distinction is necessary because the sheared flow amounts to a vorticity of 
$\omega_{\rm box} = - 3\varOmega_K/2$. The total vorticity is 
$\omega_T = \omega_V + \omega_{\rm box}$, where $\omega_V$ is the 
vortex's intrinsic vorticity. By isolating $\varOmega_V\delta$ and dividing 
by $\varOmega_K$, we have

\begin{equation}
  \Ro = \frac{\omega_V}{2\varOmega_K} - \frac{3}{4}. 
\end{equation}In the absence of a vortex, the Rossby number is 
still finite, $\Ro=-3/4$, because of the vorticity of the shear flow. 
In this limit, \eq{eq:criterion} becomes 
\[0<k<\sqrt{3}\varOmega_K/\va\,,\] which is the criterion for the MRI 
(Balbus \& Hawley 1991). As we shall see, the growth rate in this limit 
also matches that of the MRI. This suggests that the MRI is a particular 
case of the magneto-elliptic instability.

Since we measured $\omega_V/\varOmega_K\approx-0.6$, the Rossby number is  
approximately $\Ro\approx -1$. These results are compatible with the Kida 
solution (Kida 1981)

\begin{equation}
  \varOmega_V = \frac{3\varOmega_K}{2(\chi-1)},
\end{equation}from which we derive

\begin{equation}
  \omega_V = -\frac{3\varOmega_K}{2}\frac{\chi+1}{\chi(\chi-1)} 
\end{equation}and thus

\begin{equation}
  \Ro = -\frac{3(\chi^2+1)}{4\chi(\chi-1)}.
\end{equation} For $\chi$=4, the expressions above yield 
$\omega_V/\varOmega_K=-5/8=-0.625$, which matches well the 
vorticity plateau measured in \fig{fig:vortex-shear}, and Rossby 
number $\Ro=-17/16\approx-1$. Since the Rossby number is 
$\Ro\approx-1$, \eq{eq:criterion} implies that the 
horizontal instability in the vortex is present when $q \apprle 2$. 

In dimensionless units, we use $L_z=0.2$, and a resolution of $N_z$=128 points 
in the $z$-direction, so the wavenumbers present in the box are 
$k_0 < k < k_{\rm Ny}$, where $k_0 = 2\pi/L_z = 31$ is the largest scale, 
and $k_{\rm Ny} = \pi/\Delta{z} = 2011$ is the Nyquist scale. 
The {\sc Pencil Code} needs eight points to resolve a wavelength without significant 
numerical dissipation, so for practical purposes, the maximum 
wavenumber of the inertial range is $k_{\rm Ny}/4 = 503$. Also 
in dimensionless units, $\mu_0$=$\rho_0$=$\varOmega_K$=1, so for 
$B_0$=$\va$=$\xtimes{5}{-3}$, the condition posed by \eq{eq:criterion}
is $k < 400$, well within the range captured by our box.

As for growth rates, 
Mizerski \& Bajer (2009) do not unveil a simple expression. The solution has 
to be computed numerically{\footnote{A script to calculate 
the growth rates was kindly provided by K. Mizerski.}}. The growth 
rate is a function of $\chi$, $q$, and the angle $\theta$ between the $z$-axis and 
the wavevector of the perturbation. Technically, the Rossby number is also a 
free parameter, but the Kida solution ties the Rossby number to the aspect ratio. 
We present the growth rates for $\chi=4$ in the $q$-$\theta$ plane in the 
left panel of \fig{fig:growth-rates}.

It is seen that the most unstable modes are those of the horizontal 
instability ($\theta=0$, or $k_z$ modes). The right panel 
of \fig{fig:growth-rates} shows the growth rates of these modes for a 
series of aspect ratios. For $\chi=2$ and $\chi=3$, we are in the range of 
existence of the classical (hydro) horizontal instability, noted by the 
fact that fast exponential growth exists for $\va=0$. For $\chi=4$ onwards, 
the instability does not exist or is too weak in the hydro regime, and the 
most unstable wavelength is found in the vicinity of $q$=1. 
Although a critical wavelength exists for $k_z$ modes, 3D
resonant instability exists for an unbounded range of 
wavenumbers, albeit with slower growth rates. In the next sections, unless 
otherwise stated, whenever we mention ``magneto-elliptic instability'' 
we mean the horizontal, nonresonant, magneto-elliptic modes.

We also calculate the growth rate in the limit of pure shear flow, which 
we approximate numerically by $\chi=100$.  In this case, there are no 
3D unstable modes, since there is no finite vortex turnover 
frequency to establish resonances. The only instability present is 
horizontal ($k_z$), which we also show in the right panel of 
\fig{fig:growth-rates}. The critical wavelength is $q\approx \sqrt{3}$, most 
unstable wavelength $q\approx 1$, with growth rate 
$\sigma\approx0.75\varOmega_K$. One immediately recognizes that these 
properties are the properties of the MRI. That the MRI is a limiting case 
of the magneto-elliptic instability makes 
sense because \eq{eq:streamlines} with the Kida solution reduce to a Keplerian 
sheared flow when $\chi$ tends to infinity. Physically, the destabilization of 
$k_z$ modes of the elliptic and magneto-elliptic instability mean exponential 
drift of epicyclic disturbances. The equivalent of epicyclic disturbances for 
$\chi\rightarrow\infty$ are radial perturbations in a sheared flow. The 
magneto-rotational and magneto-elliptic are in essence the same instability. 
Another way of seeing the common ground between the instabilities is by 
realizing that, although constant angular momentum means rigid rotation in 
circular streamlines, it does not mean so when it comes to elliptical 
streamlines. Figure~\ref{fig:circle-ellipse} illustrates this point. The 
length of a line connecting two points is conserved in uniform circular 
motion, but not in uniform elliptical motion{\footnote{As pointed out 
by the referee, this is clearly seen when one writes the shear stress 
$S_{\rm sh}$=$\partial_x{u_y}+\partial_y{u_x}$ and substitutes the Kida solution. It yields
$S_{\rm sh}$=$-\Omega_V(\chi^2-1)/\chi$, which is only zero for $\chi$=1.}}. 
In other words, uniform elliptical motion contains shear. A magnetic field connecting the 
two points will resist that shear, leading to instability depending on the field strength. 

Judging from the \fig{fig:growth-rates}, 
the growth rates of the magneto-elliptic instability at the Balbus-Hawley wavelength
$q=1$ seem to be well-reproduced by a fit 

\begin{equation}
  \sigma_{\rm BH} \approx 0.75\varOmega_K \, \left(\frac{\chi+1}{\chi}\right),
\end{equation}i.e., scaled by a factor $(\chi+1)/\chi$ with respect  
to those of the MRI. We hereafter refer to this $\chi=100 \gg 1$ curve 
as the MRI limit. 

\subsection{Isolating the vortex magnetic action}

As seen in \fig{fig:growth-rates}, the wavelength range of the 
magneto-rotational and (horizontal) magneto-elliptic instabilities 
are almost the same for the aspect ratio of 
interest, leaving only a narrow range where one instability is captured 
but not the other. However, the growth rates differ, and we can explore 
this fact. The maximum growth rate for $\chi=4$ is 
$\sigma\approx0.95\varOmega_K$. While the MRI is amplified a millionfold 
in three orbits, the magneto-elliptic instability is amplified by more 
than a billionfold in the same time interval. We study in this section 
limiting cases 
where the instabilities do not grow as fast as in \fig{fig:evolution-oo-bb}, thus 
allowing us to better study their behavior. Because the magneto-rotational and 
magneto-elliptic instabilities will both be present in the simulations, we 
loosely refer to them collectively as ``the MHD instabilities'' or just 
``the instabilities'' in the next sections. 

\begin{figure*}
  \begin{center}
    \resizebox{.82\textwidth}{!}{\includegraphics{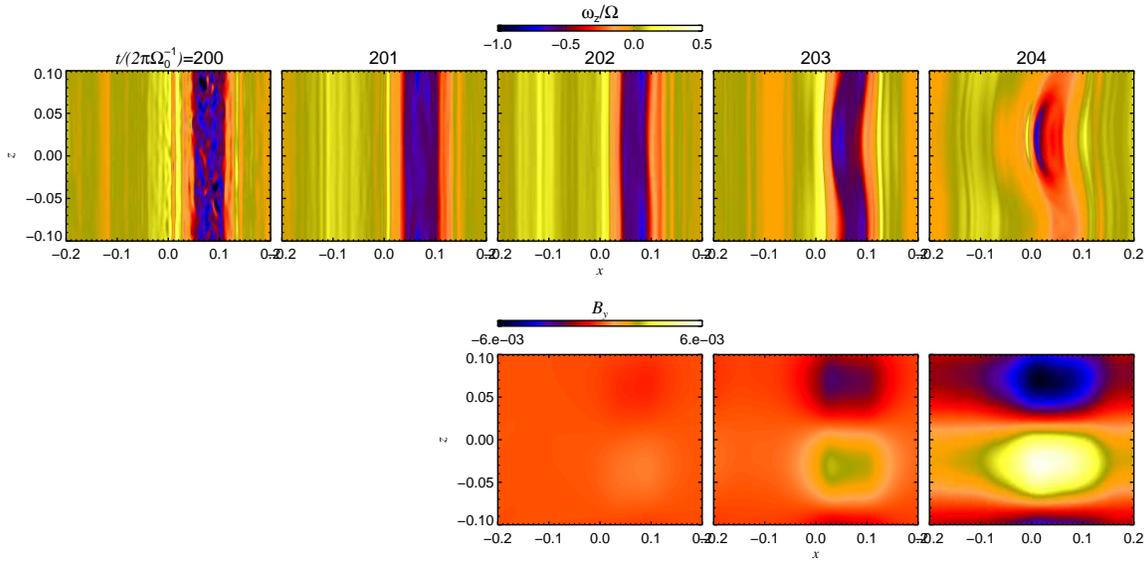}}
\end{center}
\caption[]{The effect of a strong unstable vertical magnetic field in the 
vorticity column. The field is added at t=200. At first, the effect of the 
field is to stabilize the elliptic turbulence, which is seen in the subsequent 
snapshots. The disappearance of the vortex 
at later times is caused by the development of a strong channel flow that stretches 
the column and destroys its vertical coherence. If the initial vertical field is 
stable, the strength of the channel does not grow and the vortex survives 
indefinitely.} 
\label{fig:channel-flow}
\end{figure*}

\begin{figure*}
  \begin{center}
    \resizebox{.82\textwidth}{!}{\includegraphics{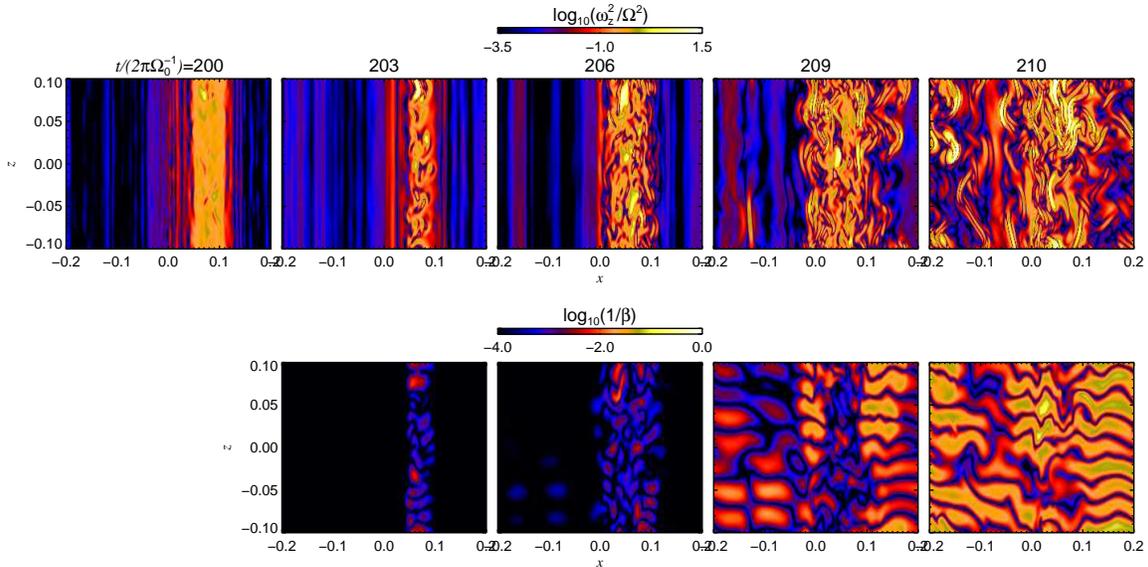}}
\end{center}
\caption[]{Time series of enstrophy and plasma beta for run R, where 
the instabilities grow at lower growth rates than in run P (\fig{fig:evolution-oo-bb}).
The magneto-elliptic instability grows faster than the MRI, which is seen as the strong 
turbulence that develops in the core, while the underlying Keplerian flow is still laminar. 
Once the MRI saturates, the strain of its turbulence destroys the vortex spatial coherence. 
It is not conclusive if the vortex would have survived the magneto-elliptic instability had the 
MRI not destroyed it first.}
\label{fig:vortex-mri}
\end{figure*}

\begin{figure*}
  \begin{center}
    \resizebox{.82\textwidth}{!}{\includegraphics{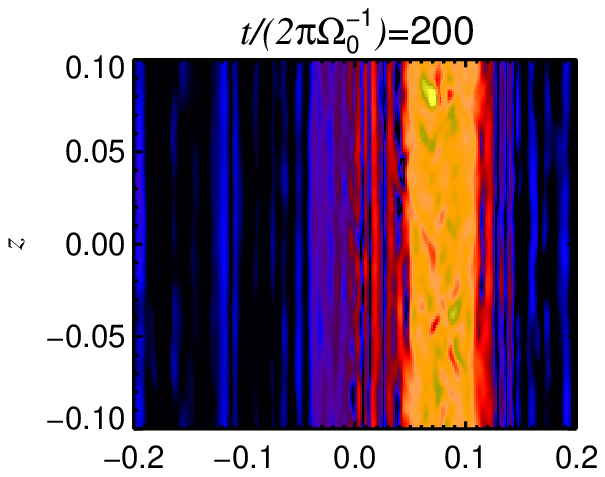}}
    \resizebox{.82\textwidth}{!}{\includegraphics{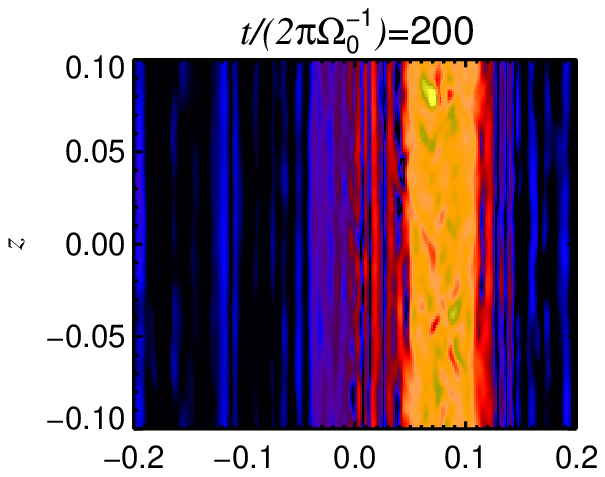}}
\end{center}
\caption[]{Time series of enstrophy and plasma beta for runs S, where 
the instabilities are quenched with resistivity. The upper panels correspond to a 
high-resistivity run, where even the longest wavelength of the box is damped. The simulation
is similar to a nonmagnetized run, the vortex surviving indefinitely. In the lower panels 
we used lower resistivity with Els\"asser number $\varLambda$=1. The longest wavelength 
of the box thus has a magnetic Reynolds number of 6, so its growth is not quenched.  
The magneto-elliptic instability grows in the vortex core in a conspicuous 
$k_z/k_0$=2 mode. Part of the field generated is diffused away due to the high resistivity. 
Channel flows eventually develop, destroying the vortex.}
\label{fig:resistivity}
\end{figure*}

\begin{figure*}
  \begin{center}
    \resizebox{.9\textwidth}{!}{\includegraphics{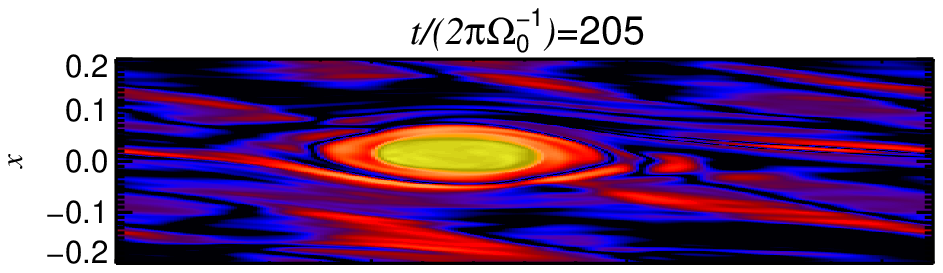}}
\end{center}
\caption[]{Time series of enstrophy and azimuthal field in the midplane for run S2, 
of moderate resistivity. By action of the magneto-elliptic instability, the field 
initially grows inside the vortex. Due to the resistivity, it then diffuses away 
from the vortex, coupling to the waves excited by it. At later times, exponential 
growth of the field is seen in the wake. The vortex itself appears unmagnetized.}
\label{fig:resistivity-xyplane}
\end{figure*}

\subsubsection{Increasing the field strength - Stabilization of elliptic instability and channel flows}
\label{sect:Increasing-the-field-strength}

We add to the box a vertical field of strength $B_z$=$\xtimes{6}{-2}$. 
Since the smallest wavenumber of the box is $k_0$=31, we have that the critical wavenumber for the 
MRI is $k/k_0 = 0.9$ and thus the box is MRI-stable. The critical wavenumber for the 
magneto-elliptic instability, according to \eq{eq:criterion}, is $k/k_0=1.13$ and thus in principle 
resolved. We aim with this to explore the window between $\sqrt{3} < q < 2\sqrt{|\Ro|}$ where 
the MRI is suppressed but not the horizontal magneto-elliptic instability.

We follow the evolution of box-average quantities in \fig{fig:magruns}. The run in question
is shown in that plot as green dot-dot-dot-dashed lines, and corresponds to run Q1 
in \Tab{table:magnetic-disk-models}.

After insertion of the field, we immediately see a decrease in the box average of the 
vertical velocities. The vertical vorticity is unchanged. Radial and azimuthal fields 
decay with the decay of the 
vertical velocity. A weak vertical magnetic field of rms $\beta$=1000 is sustained. 
Even though the analysis provides an elliptical wavelength that 
is shorter than the box
length, we do not seem to witness a magneto-elliptic instability in operation. In fact, 
we are in the range of stable Rossby numbers, evidence of which is that the 
elliptic instability was stabilized after inserting the field. This is not 
really worrisome considering that the derived critical wavenumber was so close to $k_0$, 
and we made some approximations. It is curious, though, that we do not see growth in the unstable resonant modes. For wavelengths emcompassing the vortex core 
($\lambda_y$=$2H$ or $\lambda_x$=$H/2$; each of them well-resolved with 32 points), the 
maximum growth rate is at the vicinity of $\sigma=0.33\Omega_K$, yielding a millionfold 
amplification in less than 7 orbits. We ran this simulation for 30 orbits after inserting the field. At present, we can offer no explanation as to why these modes did not become unstable.

We also test a run with a slightly less strong field, $B_z$=$\xtimes{3.75}{-2}$ (run Q2). 
In that case, the magneto-rotational and magneto-elliptic instabilities have 
critical wavelengths of $k/k_0=1.47$ and $k/k_0 = 1.78$, respectively, so
both instabilities ought to be resolved. The largest scale of the box corresponds 
to $q=1.18$, close to the maximum growth rate of both instabilities. 
The magneto-elliptic instability has a faster growth rate, so it should be seen first.

What we witness is quite revealing. The vortex is destroyed in 4 orbits, while the MRI 
is still growing in the box. A growth of magnetic energy occurs within the vortex
at a very fast pace. The vortex is destroyed when still in the phase 
of linear growth of the instability, owing to the development of a conspicuous 
and strong channel flow (\fig{fig:channel-flow}). Because the flow in 
different layers occurs in different directions, the vortex is stretched apart 
and loses its vertical coherence. 

We notice that, prior to the excitation of the channel flow, the elliptical instability 
in the core was suppressed, which is also obvious from comparing the snapshots at 
$t$=200 and 201 orbits at \fig{fig:channel-flow}.

The run is also shown in \fig{fig:spacetime} (run Q). We see that the growth of
magnetic energy occurs earlier in the vortex when compared to the surrounding flow, 
as expected.

\subsubsection{Decreasing the field strength - Vortex MHD turbulence}
\label{sect:Decreasing-the-field-strength}

Next we checked the behavior of the flow by adding weak magnetic fields. 
The goal was to slow the MHD instabilities by not resolving their most unstable wavelengths, 
$q\approx 1$. Both instabilities thus operate at a slower pace, which results 
in stretching the time interval while one (magneto-elliptic in the vortex) is 
saturating and the other (magneto-rotational in the box) still growing.

The cell size in the $z$-direction is $\xtimes{1.6}{-3}$. We add a field of
strength $B_z$=$\xtimes{1.5}{-3}$. The Balbus-Hawley wavenumber is 
$k_{\rm BH}=667$, and thus resolved but within the viscous range. The first 
properly resolved wavenumber is $k\approx500$, which corresponds to $q\approx0.75$. 
The run is shown in \fig{fig:magruns} as dashed red line, and labeled R in 
\Tab{table:magnetic-disk-models} and \fig{fig:spacetime}.

It is seen that the MRI in the 
Keplerian flow is suppressed, yet an instability is present. We identify it
with the magneto-elliptic instability, as it coincides with the vortex 
core going unstable, as shown in the snapshots of \fig{fig:vortex-mri}.

The vortex is magneto-elliptic unstable, yet it does not seem to lose its spatial 
coherence. The magnetic field is mostly confined to the vortex, which shows as a region of high Alfv\'en speeds, 
when the surrounding Keplerian flow is still laminar. The instability is violent, making 
the vortex bulge. This is apparent in \fig{fig:vortex-mri} as the vortex seems to have 
grown radially from $t$=203 to $t$=206 orbits. During this period, however, the 
box average of kinetic energy and enstrophy are nearly constant (\fig{fig:magruns}), so 
it is not clear if this magneto-elliptic turbulence would have led to vortex destruction, or 
if it would have reached a steady state. The process just outlined is well-illustrated 
in \fig{fig:spacetime} (run R). One orbit later, the MRI starts to 
develop in the surrounding Keplerian flow (notice the difference between these time scales 
and those of \fig{fig:evolution-oo-bb}), which corresponds to the increase in box-average quantities 
in \fig{fig:magruns} at that time. No strong channel flow is excited. The level of vorticity 
due to the MRI is nonetheless bigger than that of the vortex. The latter eventually becomes 
inconspicuous in the midst of the box turbulence. 

We also tested a weaker field, of strength $B_z$=$\xtimes{6}{-4}$. The wavenumbers of the 
analysis above were then scaled by 2.5, so the first resolved wavenumber corresponds to 
$q=0.3$. In this case, no significant action was seen. After ten orbits, the intensity 
of the magnetic energy was only $\xtimes{4}{-9}$, accompanied by a merely slight increase 
in the kinetic energy of the vertical velocities ($<$$u_x^2$$>$ and $<$$u_y^2$$>$ remained unchanged). The minimum plasma beta was still as high as $\ttimes{4}$.

\subsubsection{Resistivity}
\label{sect:Resistivity}

To test the last case, we used a resistivity high enough that the longest unstable 
wavelength present in the box has a magnetic Reynolds number of unity. This 
wavelength is of course $L_z$, the vertical length of the box. The resistivity then is such that 
$\mathrm{Re}_M=L_z\va/\eta=1$; for a field of strength $B_z$=$\xtimes{5}{-3}$, this 
magnetic Reynolds number corresponds to $\eta$=$\ttimes{-3}$. This is the same 
field that was used in the fiducial MHD run (\fig{fig:evolution-oo-bb}), of 
$k_{\rm BH}=200$, so the instabilities are 
resolved in the absence of resistivity. The run is labeled S in 
\Tab{table:magnetic-disk-models}. The results are shown in the upper panels 
of \fig{fig:resistivity}. 

The simulation is not very different from a purely hydro run. The damped 
magnetic field only has a slight stabilizing effect on the elliptical instability. 
A slight amount of the kinetic energy of the core turbulence gets 
converted into magnetic energy, which then diffuses away. The vortex becomes, at later times, 
less magnetized than the surroundings.

The situation should change when the resistivity is lowered slightly, allowing some 
unstable wavelengths to have $\mathrm{Re}_M>1$, yet still quenching the most unstable 
wavelengths. For that, we set the Els\"asser number to
$\varLambda \equiv  2\pi\va^2/\varOmega_K\eta = 1$. The Els\"asser number is equivalent to 
the magnetic Reynolds number $\mathrm{Re}_M$=$LU/\eta$ taking the length $L$ as the 
MRI wavelength, and velocity $U$ as the Alfv\'en velocity. As such, it is the quantity 
governing the behavior of the MRI (e.g., Pessah 2010). Having $\varLambda$=1 corresponds 
to $\eta$=$2\pi\va^2/\varOmega_K$, or $\eta\simeq\xtimes{1.6}{-4}$ in dimensionless units. 
The magnetic Reynolds number of the longest wavelength is thus $\approx$ 6. 
The results are shown in the lower panels of \fig{fig:resistivity}.

The vertical field again has a stabilizing effect on the elliptical turbulence. This is 
seen  as a weakening of the vertical kinetic energy in \fig{fig:magruns}, which lasts for 
two orbits. The difference between this run and the more resistive one is that thanks to the 
excitation of magneto-elliptic modes, radial and azimuthal fields grow inside the 
vortex core, and a conspicuous $k/k_0$=2 vertical mode appears (lower panels of 
\fig{fig:resistivity}). The field gets looped around 
the vortex, initially making the vorticity patch a region of higher Alfv\'en speeds. 
Owing to the high resistivity, however, the field diffuses away (the time for the field to diffuse 
over a scale height is $t$=$H^2/\eta \approx$ 10 orbits). The radial field gets sheared into 
azimuthal by the Keplerian flow. After a few orbits, strong magnetic fields are 
seen in the vortex spiral waves. At later times, the exponential growth of radial 
and azimuthal fields, 
as well as the excited $z$-velocities, are seen in these waves. This process 
is illustrated in \fig{fig:resistivity-xyplane}.

A look at the induction equation illustrates the process. Under incompressibility and elliptical 
motion (\eq{eq:streamlines}), the equations for the in-plane field perturbations under the 
influence of a vertical magnetic field and resistivity are

\begin{eqnarray}
  \pderiv{B^\prime_x}{t}&=& B_z\partial_z u^\prime_x  + B^\prime_y\varOmega_V/\chi + \eta\Laplace{B^\prime_x}\\
  \pderiv{B^\prime_y}{t}&=& B_z\partial_z u^\prime_y - B^\prime_x\varOmega_V \chi + q\varOmega_0 B^\prime_x + \eta\Laplace{B^\prime_y}.
\end{eqnarray}

The first term in both equations generates field out of velocity perturbations. This is 
the only source term for the in-plane field. Under waning velocity perturbations, the generation 
of fields dies out as well. The second term is also stretching, but under the vortical motion, thereby turning radial fields into azimuthal and vice-versa, at the vortex frequency. Its effect is to 
wrap around the vortex the fields generated by the first term. The field then diffuses away due to the 
resistivity. The radial field is sheared into azimuthal because of the third term in the azimuthal field 
equation. 

In this simulation, the magnetic Reynolds number of the longest wavelength is 
$\mathrm{Re}_M=L_z\va/\eta$=6, so even though the most unstable wavelength of the 
MRI is suppressed, slower growing wavelengths are present. Since they amplify the field, 
strong channel flows eventually appear in the simulation. At 10 orbits the azimuthal field 
of the channel achieves the same strength as that of the field in the 
vortex. The box went turbulent at 15 orbits (\fig{fig:spacetime}), slightly after 
the destruction of the vortex by the 
channel flow. We note, however, that in the control run 
for the simulation in question, the MRI grew slower, only becoming noticeable 
at $\approx 20$ orbits (notice the larger range of the time axis for the control run). It 
appears that the field produced by the magneto-elliptic instability in the vortex 
and then diffused to the box led to the faster growth compared to the $\ksi=0$ control 
run.

A simulation where the Reynolds number of the longest wavelength was three also shows the 
same qualitative behavior, albeit in longer timescales. We followed a simulation of 
Reynolds number two for the same time, and no growth was seen. The timescale for 
growth in this case may be infinite (stable) or just impractically long. We conclude that 
resistivity suppresses the magneto-elliptic instability when the longest unstable wavelength 
has a magnetic Reynolds number of the order of unity, as intuitively expected.

\subsection{Constant azimuthal field and zero net flux field}
\label{sect:other-fields}

The analysis of the magneto-elliptic instability by Mizerski \& Bajer (2009) was done 
for a system thread by a uniform constant magnetic field. We seek here to establish 
the effect of a zero-net flux field. As it turns out, the vortex is quite unstable 
to such configurations as well. 

We add a field whose initial value is $B$=$B_0 \sin(k_{z0} z)\hatx$, where $k_{0z}$=$2\pi/L_z$, 
and $B_0=\ttimes{-2}$. The run is labeled U in \Tab{table:magnetic-disk-models} and 
\fig{fig:spacetime}. The most unstable wavelength for the MRI has $k=100$, hence well-resolved. 
In a barotropic box, this field led to saturated turbulence after 3 orbits. The critical 
wavelength for the magneto-elliptic instability has $k=200$, also well resolved. As shown 
in \fig{fig:spacetime}, the vortex becomes unstable well before the box turbulence starts. 
In 1 orbit after insertion of the field, the vortex column has already lost its coherence. 

As for an azimuthal field, Kerswell (1994) studied the effect of toroidal field on 
elliptical streamlines, finding only a slight stabilizing adjustment of the growth rates 
of the elliptical instability. The analysis, however, only holds for the limit of 
nearly circular streamlines ($\chi \rightarrow 1$). Given the stark difference in the 
behavior of vertical fields in different configurations, there is reason to believe the 
same should apply to azimuthal fields. We add a field $B=B_0\haty$, with $B_0=0.03$ (run 
T in \Tab{table:magnetic-disk-models}). Once again, the vortex is quickly destroyed, 
as seen in \fig{fig:spacetime}.

\section{Conclusions}
\label{sect:Conclusions}

We model for the first time the evolution of the baroclinic instability 
in 3D including compressibility and magnetic fields. We find 
that the amount of angular momentum transport due to the inertial-acoustic 
waves launched by unmagnetized vortices is at the level of $\alpha\approx\xtimes{5}{-3}$, 
positive, and compatible with the value found in 2D calculations.

When magnetic fields are included and well-coupled to the gas, an MHD instability
destroys the vortex in short timescales. We find that the 
vortices display a core of nearly uniform angular velocity, as claimed in the 
literature (e.g., Klahr \& Bodenheimer 2006),  so this instability is not the MRI. 
We identify it with the magneto-elliptic instability studied by 
Lebovitz \& Zweibel (2004) and Mizerski \& Bajer (2009). 

Though Lebovitz \& Zweibel (2004) report that the magneto-elliptic instability has 
lower growth rates than the MRI, our simulations show the vortex core going unstable 
faster than the box goes turbulent. That is because the presence of background 
Keplerian rotation allows for destabilization of $k_z$ modes (horizontal instability), 
which have higher growth rates. We also show that the stability 
criterion and growth rates for the magneto-elliptic instability derived by Mizerski 
\& Bajer (2009), when taken in the limit of infinite aspect ratio (no vortex) and with 
shear, coincide with those of the MRI. Both instabilities have a similar most unstable 
wavelength, yet the growth rates of the magneto-elliptic instability in the range of 
aspect ratios $4 < \chi < 10$ are approximately 3 times faster than for the MRI.

After the vortex is destroyed, the saturated state of the MRI+BI simulation 
resembles an MRI-only simulation. The same box-averaged values of $\alpha$, 
enstrophy, kinetic, and magnetic energies are measured in the two cases. The conclusion is that the 
background entropy gradient plays only a small role when magnetic fields are present and well-coupled to the gas. The enstrophy produced by the BI is four orders of magnitude lower than 
that produced by the MRI.

We performed a series of numerical experiments to determine the behavior of the 
magneto-elliptic instability in limiting cases. First, we increased the field so that the critical MRI wavelength 
is bigger than the box. In that case, the elliptical turbulence dies out almost immediately 
after inserting the field. We take it as evidence of the stabilizing effect 
of strong magnetic fields on the classical (hydro) elliptical instability. 
When the field is slightly decreased so that some unstable wavelengths are 
resolved, the magnetic field inside the 
vortex core grows rapidly, leading to channel flows that soon break the spatial coherence 
of the vortex column.

Second, we slow down the instabilities to better study the magneto-elliptic instability 
in isolation. Decreasing the growth rate by a factor x stretches the time period between their saturations 
by the same factor. We thus decreased the field so that the most unstable wavelength 
in the box is underresolved. In this case, we witness the development of 
magneto-elliptic turbulence in the vortex core only. This turbulence was 
violent, but it is not clear if it would have led to destruction of the 
vortex. After seven orbits, longer MRI-unstable wavelengths in the box led to turbulence. The vortex 
was destroyed by the strain of that turbulence, bulging and losing coherence, and was 
eventually lost in the turbulent vorticity field of the box. Decreasing the field further led to 
quenching of the magneto-elliptic instability as well.

Third, the instabilities were suppressed with physical resistivity, setting 
the Els\"asser number to unity. In this case, there is a slight decrease in the 
kinetic energy of the elliptic turbulence, which lasts for two orbits. Meanwhile, 
in-plane magnetic fields develop inside the vortex, loop around it, and get diffused 
away. Vortex destruction happens when longer wavelengths in the box, for which the magnetic 
Reynolds number is bigger than one, go MRI-unstable. Channel flows develop, and the vortex is 
stretched apart. By increasing the resistivity, the instabilities are quenched 
when the longest wavelength of the box has a magnetic Reynolds number $\Rey_{\rm M} \apprle 2$.

In addition to uniform vertical fields, we also performed simulations with 
azimuthal fields and vertical zero net flux fields. These different field 
configurations also led to magneto-elliptic instability in the vortex.

In view of these results, it 
is curious that the vortex seen in the zero net flux MRI simulations of 
Fromang \& Nelson (2005) is stable over hundreds of orbits, a fact 
left without explanation to date. We speculate that this may come from a 
lack of resolution in the global simulation to capture the magneto-elliptic 
modes in the core.

We conclude that the baroclinic instability is important only when magnetic fields 
are too weakly coupled to the gas. Otherwise they are destroyed by the magneto-elliptic 
instability, channel flows, or by the strain of the surrounding MRI turbulence. We 
thus underscore that 
our results fit neatly in the general picture of the layered accretion paradigm in 
protoplanetary disks. If the MRI supersedes the BI, it thus remains 
the main source of turbulence in the active zones where ionization is 
abundant. The active layers are unmodified, whereas the dead zone, if 
baroclinic unstable, is endowed with large-scale vortices and an associated weak 
but positive accretion rate of $\alpha \approx \xtimes{5}{-3}$. This 
value has to be revised by global simulations in view of the limited spatial scale of 
the shearing box. If confirmed, it might be sufficient for a steady state 
to be achieved (Terquem 2008), as long as the borders of the dead zone are stable 
(Lyra et al. 2009). It remains to be studied what the conditions are when vertical 
stratification is included, what the precise transition is between a BI 
dominated dead zone and the MRI-active radii/layers, and how the accumulation of 
solids will proceed inside elliptically turbulent vortex cores. 

\begin{acknowledgements}
Simulations were performed at the PIA cluster of the Max-Planck-Institut
f{\"u}r Astronomie. We acknowledge useful discussions with K. Mizerski,
C. McNally, J. Maron, M.-M. Mac Low, G. Lesur, and A. Johansen. The 
authors thank the anonymous referee for the many comments that helped improve 
the manuscript.
\end{acknowledgements}
\Online
\begin{appendix}

\section{Linearization of the large-scale pressure gradient.}
\label{app:linearization}

Given a pressure profile following a power law 

\begin{equation}
  \bar{p} = p_0\left(\frac{r}{R_0}\right)^{-\xi},
\end{equation}we linearize it by considering $r$=$R_0+x$, 
and $R_0 \gg x$, as below

\begin{equation}
  \bar{p} = p_0(1-\xi x/R_0). 
\end{equation} 

The total pressure includes this time-independent part, plus a local 
fluctuation $p^\prime$

\begin{eqnarray}
  p &=& \bar{p} + p^\prime \nonumber \\
    &=& p^{(0)} -  p_0\xi x/R_0, 
  \label{eq:totalpressure}
\end{eqnarray}where

\begin{equation} 
  p^{(0)} =  p_0 + p^{\prime}, 
  \label{eq:localpressure}
\end{equation}and the superscript ``(0)'' denotes the 
part of the pressure that has no information on the radial 
gradient. The pressure gradient is therefore

\begin{equation} 
  \grad{p} = -p_0\xi/R_0\hatx + \grad{p^\prime} = -p_0\xi/R_0\hatx + \grad{p^{(0)}}.
\end{equation} 

The momentum equation then becomes

\begin{equation} 
  \frac{D\v{u}}{Dt} = -\frac{1}{\rho}\grad{p} = -\frac{1}{\rho}\grad{p^{(0)}} + \frac{p_0\xi}{R_0\rho}\hatx. 
\end{equation}However, this equation would not work because it is not in equilibrium. The 
second term would continuously inject momentum in the box. This is a reflection 
of the fact that the pressure gradient modifies the effective gravity, and the rotation curve 
accordingly. In a global disk, the disk settles in sub-Keplerian centrifugal equilibrium. We 
add to the equation a term that embodies this equilibrium 

\begin{equation} 
  \frac{D\v{u}}{Dt} = -\frac{1}{\rho}\grad{p} = -\frac{1}{\rho}\grad{p^{(0)}} + 
\left(\frac{1}{\rho}-\frac{1}{\rho_0}\right)\frac{p_0\xi}{R_0}\hatx. 
\end{equation}

The superscript ``(0)'' is dropped in \eq{eq:navier-stokes}. The same procedure applies 
to the energy equation, albeit with a small caveat, because the energy 
equation includes not only the pressure gradient but also the pressure itself. In 
adiabatic form, the energy equation is 

\begin{equation}
  \frac{\partial E}{\partial t} + \left(\v{u}\cdot\del\right) E = -\frac{\gamma}{(\gamma-1)}p\Div{\v{u}}, 
\end{equation}where $E=\cv\rho T$ is the internal energy. The term on the right-hand side 
is the pdV work. The modifications to the energy equation come from this term and from the 
radial advection term. Since $E=p/(\gamma-1)$, and $p$ is given by \eq{eq:totalpressure}, 
we have

\[\partial_x E = \partial_xE^{(0)} - p_0\ksi/[R_0(\gamma-1)],\]so

\begin{eqnarray}
  \frac{\partial E^{(0)}}{\partial t} &+& \left(\v{u}\cdot\del\right) E^{(0)} = \\\nonumber
  &&-\frac{\gamma}{(\gamma-1)}\left[p^{(0)} - p_0\ksi\frac{x}{R_0}\right]\Div{\v{u}} + \frac{u_x p_0\ksi}{R_0(\gamma-1)}.
  \label{eq:energy-equation}
\end{eqnarray}

We now {\it drop} the $x$-dependent term in the pdV work. The main reason 
is because the term is not periodic, therefore its inclusion in the 
shearing box can potentially lead to long-term effects in the simulation, e.g., 
uneven heating due to the dissipation of waves. It also creates a boundary 
in the box, and waves would be subject to refraction and reflection upon 
reaching it.

Although the discarding is not physically motivated, the loss is not as crucial 
as it may seem at first. This is because the term 
is not important for the instability cycle; that is, the pdV is not acting on 
the buoyancy itself. Clearly a vortical flow in geostrophic balance is 
divergenceless, so the pdV term is zero anyway. Thus, neglecting the x-dependent 
term is not affecting the baroclinic instability. If included, the term 
is only affecting the wave pattern, and only slightly. 
Waves will be subject to more heating (cooling) upon compression (expansion) 
on one side of the box more than the other. In fact, this could also be expressed 
by a radially varying adiabatic coefficient $\gamma$. A test simulation was run 
with the non-periodic term included, in order to assess the impact of 
discarding it. It was seen that the term effectively creates, 
over a long time, a small entropy jump in the box. Nonetheless, 
the development of the baroclinic instability 
progressed unhindered, leading to no significant different results to vortex 
formation. Even the large-scale sound waves have the same statistical properties. 
Growth rates and saturation levels of the turbulence are indistinguishable 
from those attained in the fiducial run. 

We are of course losing consistency for dropping a term from the equations. The 
consistent ways of modeling the problem are either by using the Boussinesq 
approximation (as in Lesur \& Papaloizou 2010), which filters acoustics 
out and eliminates the pdV term, or by means of global disk calculations, where 
the radial dependencies are only given in the initial condition, and are free 
to evolve in time. This is admittedly a limitation of the model, but the 
benefit of the high resolution that can be achieved in the local box, 
at the expense of a term 
that is not important for the instability cycle, was judged worth the 
loss in consistency. 

The simplified energy equation reads as

\begin{equation}
  \frac{\partial E^{(0)}}{\partial t} + \left(\v{u}\cdot\del\right) E^{(0)} = -\frac{\gamma}{(\gamma-1)} p^{(0)} \Div{\v{u}} + \frac{u_x p_0\ksi}{R_0(\gamma-1)};
  \label{eq:energy-equation}
\end{equation}i.e., the box has the same temperature all over. The dependency 
on $u_x$ provides the heating/cooling that a gas parcel would experience in 
a global model. Because of it, a gas parcel heats up when climbing the temperature 
gradient, and cools down when descending it.

From the definition of entropy (\eq{eq:entropy}), we write 

\begin{equation}
  s/\cv = \ln{\frac{E}{E_0}} - \gamma \ln{\frac{\rho}{\rho_0}}
\end{equation}and taking the derivative, 

\begin{equation}
  \frac{1}{\cv}\frac{Ds}{Dt}  = \frac{1}{E}\frac{DE}{Dt} + \gamma \Div{\v{u}}
\end{equation}where we made use of the continuity equation. Multiplying the whole 
equation by $E$, we have 

\begin{equation}
  \rho T\frac{Ds}{Dt}  = \frac{DE}{Dt} + \frac{\gamma}{(\gamma-1)} p\Div{\v{u}},
  \label{eq:entropy-convert}
\end{equation}so the pdV term cancels when converting the energy equation 
into an equation for entropy. We substitute \eq{eq:energy-equation} into 
\eq{eq:entropy-convert} to find 

\begin{equation}
  \frac{Ds^{(0)}}{Dt} =  \frac{1}{\rho T^{(0)}}\left\{\frac{u_x p_0\ksi}{R_0(\gamma-1)}\right\}.
\end{equation}The superscripts ``$(0)$'' are dropped in \eq{eq:entropy}.
 
\end{appendix}

\begin{appendix}

\section{Testing for aliasing}
\label{app:aliasing}

One critical task to perform before any attempt to quantify baroclinic growth 
of vorticity is to test how well the code reproduces the analytical results 
of shear wave theory. Compressible and incompressible modes have well-defined 
analytical solutions that may be used to assess the presence and 
quantify the amount of aliasing introduced by the scheme. 

Aliasing is a feature of finite-difference codes, which occurs when a 
shear-wave swings from leading ($k_x<0$) to trailing ($k_x>0$). As the radial 
wavelength of the wave approaches zero, it becomes shorter than 
the width of a grid cell. The signal is lost on the Nyquist scale, 
and information is lost owing to the phase degeneracy that is established. There are an infinite 
number of possible sinusoids of varying amplitude and phase that are 
solutions (all ``aliases'' of the correct solution), so spurious power 
can be introduced in the wave. This extra power is then 
transferred from the trailing to the leading wave, which will again 
swing-amplify. 

Because of this, it is possible that aliasing by itself may lead to 
spurious vortex growth. In 2D, the energy spuriously generated 
at the aliased swing-amplification has no option but to undergo an inverse 
cascade, the end of which is coherent vortices. This is a particular 
danger for {\sc Pencil}, because the code is both finite-difference and high-order. 
The high-order nature is in most cases a plus, of course, since it 
leads to little overall numerical dissipation. In this case, however, it 
means that the spuriously added power will not be discarded. Lower order 
codes can be diffusive enough that the energy introduced by aliasing may be 
immediately dissipated. Indeed, Shen et al. (2006) highlight that they do 
not see any aliasing happening, and suggest that this is due to the high degree 
of numerical diffusivity in their code. This is of course a case of two 
negative features canceling each other. 

\begin{figure}
  \begin{center}
    \resizebox{.95\hfwidth}{!}{\includegraphics{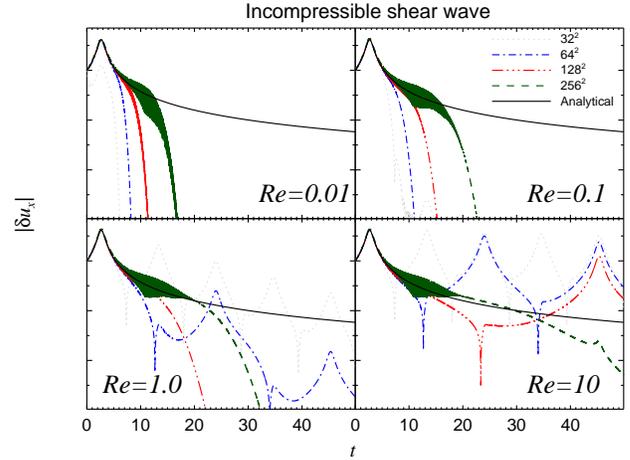}}
\end{center}
\caption[]{Testing the numerical scheme for aliasing, a feature of 
finite difference methods, which spuriously increases the energy of a 
wave that swings from leading to trailing. At the resolutions and 
mesh-Reynolds numbers we use, this effect is successfully suppressed. The wave
has a wavenumber $k_{x,0}$=$-16\pi/L_x$, which means 4, 8, 16, and 32 points per 
wavelength at resolutions 32, 64, 128, and 256, respectively. The wiggling is 
due to excitation of compressible modes not present in the analytical solution.}
\label{fig:incompressible}
\end{figure}

We combine the best of both worlds by using hyperviscosity. It makes the code 
dissipative only where it is needed (on the grid scale), with the extra benefit 
of being able to control how much dissipation is added to the solution. We note 
that Oishi \& Mac Low (2009), also using the {\sc Pencil code}, tested the numerical 
evolution of incompressible 2D as well as compressive 2D and 3D disturbances 
against their analytical solutions. They found aliasing unimportant when using 
hyperviscosity. We repeat here the incompressible test (showed in \fig{fig:incompressible}), 
and refer to Oishi \& Mac Low (2009) and Johansen et al. (2009) for the other tests. 

The analytical solution of the incompressible shear wave is (Johnson \& Gammie 2005)

\begin{equation}
  \delta u_x = \delta u_{x,0}\frac{k_0^2}{k^2}
\end{equation}where $k^2$=$k_x^2+k_y^2$, $k_y$$\equiv$$\cte$, and 

\begin{equation}
  k_x=k_x(t)=k_{x,0} + q\varOmega_0k_y t.
\end{equation}The condition of incompressibility $k_i\delta{u_i}=0$ thus dictates that $\delta u_y$=$-k_x\delta u_{x}/k_y$. The solenoidality of the wave is guaranteed by initializing the velocity field through a streamfunction 
$\v{u}$=$\curl{(\psi\hatz)}$, with $\psi$=$Ak_y^{-1}\sin\left(k_xx+k_yy\right)$. We use the setup of Shen et al. (2006), $k_{x,0}$=$-16\pi/L_x$, $k_y$=$4\pi/L_y$, with $L_x$=$L_y$=$0.5$, $A$=$\ttimes{-4}$, $c_s$=$\varOmega$=$\rho_0$=1, $\gamma$=7/5, and $q$=3/2.

We follow the evolution of the $x$-velocity in a point of the grid, 
and plot the result in \fig{fig:incompressible}, checking for differences 
due to resolution and initial mesh Reynolds number Re, defined by the 
hyperviscosity coefficient

\begin{equation}
  {\rm Re} = \frac{\delta{u_{x,0}} \,\, dx^5}{\nu^{(3)}}
\end{equation}

We see that no aliasing is detected at Re=0.01. The signal is 
viscously damped before it 
can be swing-amplified. At Re=0.1, aliasing occurs at resolution 32$^2$. 
However, the added signal suffers a severe (hyper-)viscous damping, so that 
the next swing amplification has little left to work with. When increasing the 
Reynolds number to Re=1, aliasing now happens also for resolution 64$^2$, 
but the solution is decaying. Aliasing is not detected at any Reynolds number 
for resolution 256$^2$. A run with Re=$\infty$ for resolution 32$^2$ 
(not showed) kept periodic intervals of aliasing without dissipation but 
without net growth, at least until $t$=1000.

The panels in \fig{fig:incompressible} are for $\ksi$=0. When we 
use non-zero $\ksi$, the aliased solutions show an increase in 
amplitude. The same does not happen for the solutions at low Reynolds 
numbers. The wiggling seen at higher resolution in \fig{fig:incompressible}
is accompanied by changes in the density, which leads us to conclude that 
they come from the excitation of compressible modes not present in the 
analytical solution. We use a resolution of 256$^2$ for 2D runs. At 
this resolution, we rest assured that aliasing is not happening for 
the considered Reynolds numbers. 

\end{appendix}

\end{document}